\documentclass[a4paper,11pt]{article}
\usepackage{amsmath}
\usepackage{amssymb}
\usepackage{hyperref}
\hypersetup{colorlinks=true,allcolors=blue}

\usepackage{multirow}
\usepackage{accents}
\makeatletter
\def\widebar{\accentset{{\cc@style\underline{\mskip10mu}}}}
\makeatother

\numberwithin{equation}{section}
\numberwithin{figure}{section}
\numberwithin{table}{section}

\title{}
\author{}
\date{}
\def\papertitlepage{\baselineskip 3.5ex\thispagestyle{empty}}
\def\Title#1{\baselineskip 1cm \vspace{1.5cm}%
  \begin{center}{\Large\bfseries#1}\end{center}\vspace{0.5cm}}
\def\Authors#1{\begin{center}\renewcommand{\thefootnote}{\fnsymbol{footnote}}{\it #1}\end{center}}
\def\Abstract{\vspace{1.0cm}%
  \begin{center}{\large\bf Abstract}\end{center}}

\def\mathd{\mathrm{d}}
\def\mathe{\mathrm{e}}
\def\mathi{\mathrm{i}}

\def\mathcomma{,}
\def\mathperiod{.}

\def\str{\mathop{\textrm{str}}\nolimits}

  \def\calC{{\cal C}}
  
 \def\calH{{\cal H}} 
  \def\calL{{\cal L}}
 \def\calN{{\cal N}}

\makeatletter
\def\qopname@#1{\mathop{\fam\z@#1}\nolimits@}
\def\sec{\qopname@{sec}}
\def\csc{\qopname@{csc}}
\def\sech{\qopname@{sech}}
\def\csch{\qopname@{csch}}
\makeatother

\def\QB{Q_{\text{B}}}

\begin{document}
{\papertitlepage
\hbox{ }\vspace*{0cm}
{\hfill\begin{minipage}{4.2cm}
DESY-12-110\par\noindent
June, 2012
\end{minipage}}
\Title{On semiclassical analysis of pure spinor superstring\\ in an $AdS_{5}\times S^{5}$ background}
\Authors{{\sc Yuri Aisaka${}^{1,2}$\footnote{\tt yuri.aisaka@desy.de}},
{\sc L. Ibiapina Bevilaqua${}^{3}$\footnote{\tt leandro@ect.ufrn.br}}
and\/ {\sc Brenno C.\ Vallilo${}^{4}$\footnote{\tt vallilo@gmail.com}}
\\
${}^1$DESY Theory Group,\\[-2ex]
Notkestra\ss{}e 85, D-22603 Hamburg, Germany
\\
${}^2$Instituto de F\'{i}sica Te\'{o}rica,
S\~{a}o Paulo State University, \\[-2ex]
Rua Dr.\!\! Bento Teobaldo Ferraz 271,
01140-070 S\~{a}o Paulo, Brazil
\\
${}^3$Escola de Ci\^{e}ncias e Tecnologia, Universidade Federal do Rio Grande do Norte,\\[-2ex]
Caixa Postal 1524, 59072-970, Natal, RN, Brazil
\\
${}^{4}$Departamento de Ciencias F\'{i}sicas, Facultad de Ciencias Exactas, Universidad Andres Bello,\\[-2ex]
Republica 220, Santiago, Chile}
} 
\vskip-\baselineskip
{\baselineskip .5cm
\Abstract
Relation between semiclassical analyses of Green-Schwarz
and pure spinor formalisms
in an \(AdS_{5}\times S^{5}\) background is clarified.
It is shown that
the two formalisms have identical semiclassical partition functions
for a simple family of classical solutions.
It is also shown that, when the classical string is furthermore rigid,
this in turn implies that the two formalisms predict
the same one-loop corrections to spacetime energies.}

\newpage
\tableofcontents
\newpage
\setcounter{footnote}{0}
\section{Introduction}

Over the last decade,
the semiclassical study of string theory
in an \(AdS_{5}\times S^{5}\) background
has been a central tool for
exploring the AdS/CFT correspondence~\cite{Maldacena:1997re,Gubser:1998bc,Witten:1998qj}
beyond a supergravity approximation.
To date, an enormous amount of works has been done
extending the basic picture laid in~\cite{Berenstein:2002jq,Gubser:2002tv,Frolov:2002av},
matching quantum corrections to string energies to anomalous dimensions of
gauge invariant operators in the \(\calN=4\) super Yang-Mills theory.

Since \(AdS\) geometries that appear in the AdS/CFT correspondence
are supported by Ramond-Ramond flux, it is hard to make use of
the Ramond-Neveu-Schwarz formalism.
For an \(AdS_{5}\times S^{5}\) background,
one may either use the Green-Schwarz formalism~\cite{Green:1983wt}
or Berkovits' pure spinor formalism~\cite{Berkovits:2000fe}.
However, most of the works in the area have been done only in the former.
This is a pity because the pure spinor formalism has
many aspects that are simpler than the Green-Schwarz formalism,
and is potentially more powerful
especially if one wants more than the fluctuation spectrum around
a given classical solution.

The purpose of this article is to provide support for an equivalence
of the Green-Schwarz and pure spinor formalisms at a semiclassical level.
Using the pure spinor formalism
we perform a semiclassical analysis around a simple family of classical solutions
in an \(AdS_{5}\times S^{5}\) background
and show that the formalism reproduces
the one-loop anomalous dimensions known from the Green-Schwarz formalism.
It would be useful to exploit integrability methods for a more systematic comparison,
but in this article we stick to a down-to-earth explicit comparison.

In the rest of this introduction,
we would like to put our study into context
by briefly summarizing what has been known about the pure spinor formalism.
For a more complete list,
we refer the reader to a recent review~\cite{Mazzucato:2011jt}.

Pure spinor formalism in a flat background is defined as
a worldsheet conformal field theory with a BRST symmetry
and it allows one to quantize a string in a super-Poincar\'{e} covariant manner.
Its basics and validity have been established quite adequately.
The formalism reproduces the superstring spectrum correctly~\cite{Berkovits:2000nn}\cite{Aisaka:2008vw},
and is capable of computing tree and multi-loop amplitudes in a covariant manner~\cite{Berkovits:2000fe,Berkovits:2004px}.
There remains some subtleties at three-loops and higher~\cite{Berkovits:2006vi},
but the formalism has been very successful going far beyond (e.g.~\cite{Mafra:2011nv,Mafra:2012kh})
what have been done in other formalisms.
Also, in a flat background, it is known how the BRST symmetry of the formalism
arises from the classical Green-Schwarz action~\cite{Berkovits:2004tw}\cite{Aisaka:2005vn}.

In generic supergravity backgrounds,
both Green-Schwarz and pure spinor formalisms can be used to
describe a string at a classical level.
Equations of motion for the background fields are implied by the kappa symmetry~\cite{Siegel:1983hh}
in the former~(e.g.~\cite{Grisaru:1985fv}\cite{Witten:1985nt})
and by the BRST symmetry in the latter~\cite{Berkovits:2001ue}.
Preservation of these symmetries in worldsheet perturbation theories are expected to
characterize stringy \(\alpha'\) corrections to the background equations of motion.
However, kappa symmetry is a complicated gauge symmetry
and it is difficult to discuss them quantum mechanically.
In pure spinor formalism, kappa symmetry is replaced by
a BRST symmetry and it is straightforward to identify the conditions
for conservation and nilpotency of the BRST charge at a quantum level~\cite{Berkovits:2001ue}.
By exploiting this simplicity,
one-loop conformal invariance in generic supergravity backgrounds
has been shown in~\cite{Chandia:2003hn,Bedoya:2006ic}.

Specializing to an \(AdS_{5}\times S^{5}\) background,
a Green-Schwarz action with kappa symmetry
was constructed explicitly as a supercoset model by Metsaev and Tseytlin~\cite{Metsaev:1998it}.
The key to their construction was that the \(AdS_{5}\times S^{5}\) space
can be realized as the bosonic body
of a supercoset \(PSU(2,2|4)/(SO(4,1)\times SO(5))\)
with \(32\) fermionic directions.
The supercoset has a \(\mathbb{Z}_{4}\)-structure
(a natural extension of the notion of the symmetric coset space)
which makes it possible to rewrite the Metsaev-Tseytlin action
as a bilinear form of currents~\cite{Berkovits:1999zq}.
A classical action for the pure spinor formalism
can be explicitly written down by applying the same technique
and by introducing pure spinor variables adopted to \(AdS_{5}\times S^{5}\)~\cite{Berkovits:2000fe}.
Presumably, the pure spinor action can be
understood as a BRST reformulation of the Metsaev-Tseytlin action
but to date the expectation has not been shown explicitly.
Although these actions are constructed from currents on a group manifold,
these currents are not holomorphic.
Therefore, unlike the Wess-Zumino-Witten models,
it is not known how to solve the models based on symmetry principles.
On the other hand, both models are known to possess
an integrable structure~\cite{Bena:2003wd}\cite{Vallilo:2003nx}
and one may hope to eventually solve these models
by combining integrability and conformal field theory techniques.

Although exact quantizations of
Green-Schwarz and pure spinor superstrings in the \(AdS_{5}\times S^{5}\)
backgrounds are not within a reach at the moment,
there are no problems in performing
classical and semiclassical analyses.
In the Green-Schwarz formalism,
basics of semiclassical analysis (in particular subtleties arising from
gauge fixing Virasoro and kappa symmetries)
have been clarified in~\cite{Drukker:2000ep}
and concrete analyses around very many classical solutions
have been performed, providing strong supports in favour of the AdS/CFT conjecture.
In the pure spinor formalism,
there are no complicated gauge symmetries to be fixed and
the semiclassical analysis is straightforward.
One-loop conformal invariance in the \(AdS_{5}\times S^{5}\)
background has been shown in~\cite{Vallilo:2002mh}
and later extended to an all-loop proof~\cite{Berkovits:2004xu}.
Although the pure spinor formalism
has not been used much for computing concrete quantities in the AdS/CFT context,
it has been used in~\cite{Vallilo:2011fj} to compute the anomalous dimensions of
the Konishi multiplet at strong coupling,
and the result of~\cite{Vallilo:2011fj} is in accord with the ones predicted
from the Green-Schwarz formalism~\cite{Roiban:2011fe}
and integrability techniques~\cite{Gromov:2011de}.

So, all in all, parallel developments have been made in
the Green-Schwarz and pure spinor formalisms,
but it has never been clarified why or how
the two are equivalent at a (semi)classical level.
It is this relation of the two formalisms
we wish to address in this article.

\bigskip
The plan of this article is as follows.
In section~\ref{sec:classicalPS}
we review the classical mechanics of the pure spinor formalism in an \(AdS_{5}\times S^{5}\) background.
Section~\ref{sec:semiclassicalPS} contains the body of the article.
After a general discussion on semiclassical analyses in the pure spinor formalism,
we introduce a simple family of classical solutions and show that
one-loop corrections to spacetime energies are related to
the expectation values of the fluctuation Hamiltonians on the worldsheet.
We then compare the one-loop partition functions in the Green-Schwarz
and pure spinor formalisms and argue that they agree.
We conclude in section~\ref{sec:conclusion}
and point out some future directions.
An appendix is added to summarize our notation and conventions.

\section{Classical pure spinor superstring in \texorpdfstring{$AdS_{5}\times S^{5}$}{AdS5 x S5} background}
\label{sec:classicalPS}

We start with a brief review of the pure spinor formalism in an \(AdS_{5}\times S^{5}\) background,
with some emphases on comparison with the Green-Schwarz formalism.
To motivate the definition of the pure spinor superstring action in the \(AdS_{5}\times S^{5}\) background,
we start from an explanation of the pure spinor formalism
in trivial and generic supergravity backgrounds.

\subsection{Trivial background}
\label{subsec:revflat}
In contrast to conventional approaches to string theory,
the pure spinor formalism in a trivial background
starts off by postulating a quadratic worldsheet action
with a BRST symmetry~\cite{Berkovits:2000fe}.
For type II superstring the action is given as\footnote{See appendix~\ref{sec:appendix} for a summary of the notation.}
\begin{align}
\label{eqn:flataction}
S_{\text{flat}}
&= {1\over\pi\alpha'}\int\mathd^{2}z\Bigl({1\over2}\partial x^{a}\widebar{\partial} x_{a}
{} + p_{\alpha}\widebar{\partial}\theta^{\alpha}
{} + \widehat{p}_{\hat{\alpha}}\partial\widehat{\theta}^{\hat{\alpha}}
{} - w_{\alpha}\widebar{\partial}\lambda^{\alpha}
{} - \widehat{w}_{\hat{\alpha}}\partial\widehat{\lambda}^{\hat{\alpha}}\Bigr)
\end{align}
where $(x^{a},\theta^{\alpha},\widehat{\theta}^{\hat{\alpha}})$ ($a=0,\dotsc,9$; $\alpha,\hat{\alpha}=1,\dotsc,16$)
are the standard type II superspace variables,
$(p_{\alpha},\widehat{p}_{\hat{\alpha}})$ are conjugate momenta of $(\theta^{\alpha},\widehat{\theta}^{\hat{\alpha}})$,
and the rest are ``ghost'' variables consisting of
pure spinors $(\lambda^{\alpha},\widehat{\lambda}^{\hat{\alpha}})$ and their conjugates
$(w_{\alpha},\widehat{w}_{\hat{\alpha}})$.
As can be seen from the action,
\((p_{\alpha},\theta^{\alpha},w_{\alpha},\lambda^{\alpha})\) are left moving (holomorphic)
and \((\widehat{p}_{\hat{\alpha}},\widehat{\theta}^{\hat{\alpha}},\widehat{w}_{\hat{\alpha}},\widehat{\lambda}^{\hat{\alpha}})\)
are right moving (antiholomorphic),
and \((x^{a},\theta^{\alpha},\widehat{\theta}^{\hat{\alpha}},\lambda^{\alpha},\widehat{\lambda}^{\hat{\alpha}})\)
are all understood to carry conformal weight \(0\).

The left and right moving ghosts \(\lambda^{\alpha}(z)\) and \(\widehat{\lambda}^{\hat{\alpha}}(\widebar{z})\) are subject to
quadratic ``pure spinor constraints''~\cite{Cartan}
\begin{align}
\label{eqn:ps1}
\lambda^{\alpha}\gamma^{a}_{\alpha\beta}\lambda^{\beta}(z) = 0,\quad \widehat{\lambda}^{\hat{\alpha}}\gamma^{a}_{\hat{\alpha}\hat{\beta}}\widehat{\lambda}^{\hat{\beta}}(\widebar{z}) = 0
\end{align}
and their conjugates $(w_{\alpha},\widehat{w}_{\hat{\alpha}})$ are defined only up to
``gauge transformations''
\begin{align}
\delta_{\Omega} w_{\alpha}(z) = (\gamma^{a}\lambda)_{\alpha}\Omega_{a}(z),\quad
\delta_{\Omega} \widehat{w}_{\hat{\alpha}}(\widebar{z}) = (\gamma^{a}\widehat{\lambda})_{\hat{\alpha}}\widehat{\Omega}_{a}(\widebar{z})
\mathperiod
\end{align}
The constraints of~(\ref{eqn:ps1}) seems to imply \(10\)~constraints
for each \(\lambda^{\alpha}\) and \(\widehat{\lambda}^{\hat{\alpha}}\), but actually one half of
them is ineffective and a pure spinor has \(16-5=11\) independent components.
The ghost sector therefore is a collection of
\(11\times2\) bosonic \(\beta\gamma\) systems of weight \((1,0)\)
and has \(c=22\times 2\).
Note that the value is exactly what one needs to compensate
the central charge \(c=(10-32)\times2\) from the matter sector.

Because of the non-linear nature of the constraints of~(\ref{eqn:ps1}),
the simplicity of the ghost action in~(\ref{eqn:flataction}) appears deceptive,
but there is a nice formalism called the ``theory of curved $\beta\gamma$ systems''
(or the ``theory of chiral differential operators'')
that can be used to rigorously define the first order systems
on certain non-trivial spaces such as the pure spinor cone~(\ref{eqn:ps1}).
For more on this,
we refer the reader to the literature~\cite{Malikov:1998dw}\cite{Kapustin:2005pt,Witten:2005px,Nekrasov:2005wg}\cite{Aisaka:2008vw}.

\bigskip
The other input to the formalism, the BRST operator, is given by
\begin{align}
\label{eqn:flatQB}
\QB &= Q + \widebar{Q},\quad
Q = \int\mathd z \lambda^{\alpha}d_{\alpha}(z),\quad
\widebar{Q} = \int\mathd\widebar{z} \widehat{\lambda}^{\hat{\alpha}}\widehat{d}_{\hat{\alpha}}(\widebar{z})
\end{align}
where
\begin{align}
d_{\alpha} &= p_{\alpha} + (\gamma^{a}\theta)_{\alpha}(\partial x_{a} - {1\over2}(\theta\gamma^{a}\partial\theta)),\quad
\widehat{d}_{\hat{\alpha}} = \widehat{p}_{\hat{\alpha}} + (\gamma^{a}\widehat{\theta})_{\hat{\alpha}}(\partial x_{a} - {1\over2}(\widehat{\theta}\gamma^{a}\partial\widehat{\theta}))
\end{align}
are left and right moving
supersymmetric fermionic momenta satisfying simple operator product expansions
\begin{alignat}{2}
d_{\alpha}(z)d_{\beta}(w) &= {\gamma^{a}_{\alpha\beta}\Pi_{a}(w) \over z-w},\quad& \Pi^{a} &= \partial x^{a} - \theta\gamma^{a}\partial \theta
\mathcomma\\
\widehat{d}_{\hat{\alpha}}(\widebar{z})\widehat{d}_{\hat{\beta}}(\widebar{w}) &= {\gamma^{a}_{\hat{\alpha}\hat{\beta}}\widehat{\Pi}_{a}(\widebar{w}) \over \widebar{z}-\widebar{w}},\quad&
\widehat{\Pi}^{a} &= \widebar{\partial} x^{a} - \widehat{\theta}\gamma^{a}\widebar{\partial}\widehat{\theta}
\mathperiod
\end{alignat}
Thanks to the pure spinor constraint~(\ref{eqn:ps1}),
the BRST operator \(\QB\) of~(\ref{eqn:flatQB}) is nilpotent and it makes sense to
talk of its cohomology.
\(\QB\) acts on operators via free field operator product expansions
and physical states are found as cohomologies with ghost numbers \((1,1)\),
where \(\lambda\) and \(\widehat{\lambda}\) are defined to carry ghost numbers \((1,0)\) and \((0,1)\).
Cohomologies at other ghost numbers are interpreted as
spacetime ghosts and antifields.
The cohomology has been rather thoroughly investigated
and there is no doubt that it reproduces the well-known superstring spectrum
in the trivial background.%
\footnote{To be more precise, the theory of curved \(\beta\gamma\) systems demands
that the BRST operator be supplemented by a small extra term that takes care of
fine global issues on the pure spinor space~\cite{Nekrasov:2005wg}.
This modification is crucial for defining a composite \(b\)-ghost~\cite{Berkovits:2004px}
and for correctly reproducing the higher massive spectrum~\cite{Aisaka:2008vw}.}

Of course,
there have been attempts to explain how ``natural'' the BRST structure is.
Works taking a conventional viewpoint
have explained how the BRST structure arises from
the classical Green-Schwarz superstring~\cite{Berkovits:2004tw,Aisaka:2005vn}.
In these approaches, pure spinor ``ghosts'' in the BRST operator
are literally interpreted as the BRST ghosts for the kappa symmetry of
the classical Green-Schwarz action.
Less conventional (but potentially useful)
interpretations of the BRST structure include
its relation to the so-called superembedding formalism~\cite{Tonin:2002tt},
and recent ``twistorial'' interpretation of Berkovits~\cite{Berkovits:2011gh}.

\bigskip
Note that the pure spinor formalism does not have
the reparameterization \(bc\) ghosts as fundamental fields.
However, one may define composite operators \(b(z)\) and \(\widehat{b}(\widebar{z})\) that
makes left and right moving stress tensors \(T(z)\) and \(\widebar{T}(\widebar{z})\) BRST trivial~\cite{Berkovits:2004px}:
\begin{align}
\label{eqn:flatb}
Q b(z) = T(z),\quad \widebar{Q} \widehat{b}(\widebar{z}) = \widebar{T}(\widebar{z}),\quad
Q \widehat{b}(\widebar{z}) = \widebar{Q}b(z) = 0
\mathperiod
\end{align}
Although one cannot define the \(c\) ghosts conjugate to \(b\)'s, presence of \(b\) ghosts
is just enough for defining higher-loop amplitudes~\cite{Berkovits:2004px},
Siegel gauge vertex operators~\cite{Aisaka:2009yp} etc.

At any rate, the combination of the free field action of~(\ref{eqn:flataction})
and the BRST symmetry of~(\ref{eqn:flatQB}) is arguably much simpler
than the classical Green-Schwarz formalism with the troublesome kappa symmetry,
and the pure spinor formalism has been proved very useful
for computing amplitudes in a flat spacetime~(see e.g.\ \cite{Mafra:2011nv,Mafra:2012kh} and references therein).

\subsection{Generic supergravity background}
\label{subsec:sugraBG}

Since the pure spinor formalism is super-Poincar\'{e} covariant,
it is straightforward to generalize the flat action of~(\ref{eqn:flataction})
to a non-linear sigma model describing a string propagating in
a generic supergravity background~\cite{Berkovits:2001ue}.

Linearized coupling to a supergravity background
is described by an integrated massless vertex operator.
In the pure spinor formalism, this can be constructed from
left-right products of supersymmetric currents
$(\partial\theta^{\alpha},\Pi^{a},d_{\alpha},N^{ab})$
and
$(\widebar{\partial}\widehat{\theta}^{\hat{\alpha}},\widebar{\Pi}^{a},\widehat{d}_{\hat{\alpha}},\widehat{N}^{ab})$
as
\begin{align}
\label{eqn:IIBvertex}
V &= {1\over2\pi\alpha'}\int\mathd^{2}z\Bigl(
\partial\theta^{\alpha}\widebar{\partial}\widehat{\theta}^{\hat{\beta}} A_{\alpha\hat{\beta}}
{} + \partial\theta^{\alpha} \widebar\Pi^{b} A_{\alpha b} +
\Pi^{a} \widebar{\partial}\widehat{\theta}^{\hat{\beta}} A_{a \hat{\beta}}
{} +\Pi^{a} \widebar\Pi^{b} A_{ab}  \nonumber\\
&\quad{}+ d_{\alpha} (\bar\partial\widehat{\theta}^{\hat{\beta}} E^{\alpha}_{\hat{\beta}} + \widebar\Pi^{b} E^{\alpha}_{b})
{}+ \widehat{d}_{\hat{\alpha}}(\partial\theta^{\beta} E^{\hat{\alpha}}_{\beta}+ \Pi^{b} E^{\hat{\alpha}}_{b}) \nonumber\\
&\quad{}+{1\over2}N^{ab}
{} (\widebar\partial\widehat{\theta}^{\hat{\gamma}}\Omega_{ab\hat{\gamma}} + \widebar\Pi^{c}\Omega_{abc})
{}+{1\over2}\widehat{N}^{ab} (\partial\theta^{\gamma}\widehat\Omega_{ab\gamma}+ \Pi^{c} \widehat{\Omega}_{abc}) \nonumber\\
&\quad{}+ d_{\alpha} \widehat{d}_{\hat{\beta}}P^{\alpha\hat{\beta}} + N^{ab} \hat{d}_{\hat{\gamma}}C_{ab}^{\hat{\gamma}}
{}+ d_{\alpha} \widehat{N}^{cd} \widehat{C}^{\alpha}_{cd}
{}+ {1\over4}N^{ab} \widehat{N}^{cd} R_{abcd}
\Bigr)
\end{align}
where
\begin{align}
\label{eqn:IIBsuperfields}
&A_{\alpha\hat{\beta}},\;
A_{\alpha b},\;
A_{a \hat{\beta}},\;
A_{ab},\;
E^{\alpha}_{\hat{\beta}},\;
E^{\alpha}_{b},\;
E^{\hat{\alpha}}_{\beta},\;
E^{\hat{\alpha}}_{b},\;
\Omega_{ab\hat{\gamma}},\;
\Omega_{abc},\;
\widehat\Omega_{ab\gamma},\;
\widehat{\Omega}_{abc}, \nonumber\\
&P^{\alpha\hat{\beta}},\;
C_{ab}^{\hat{\gamma}},\;
\widehat{C}^{\alpha}_{cd},\;
R_{abcd}
\end{align}
are superfields (functions of the zero-modes of \((x^{a},\theta^{\alpha},\widehat{\theta}^{\hat{\alpha}})\)) representing fluctuations of type IIB supergravity.
Physical state condition and gauge invariance for integrated vertex operators
are given by \(QV = \widebar{Q}V = 0\) and \(\delta_{\Lambda,\Lambda'} V = Q \Lambda + \widebar{Q}\Lambda'\)
and these indeed imply linearized equations of motion and gauge invariances for
the superfields of~(\ref{eqn:IIBsuperfields})~\cite{Berkovits:2001ue}.
For example, the superpotential \(A_{\alpha\hat{\beta}}\) of lowest dimension
is found to satisfy the correct constraints and gauge invariances
\begin{align}
(\gamma_{abcde})^{\alpha\beta}D_{\alpha}A_{\beta\hat{\beta}} &=
(\gamma_{abcde})^{\hat{\alpha}\hat{\beta}}\widehat{D}_{\hat{\alpha}}A_{\beta\hat{\beta}} = 0 \\
\delta_{\Lambda,\Lambda'} A_{\alpha\hat{\alpha}} &= D_{\alpha}\Lambda_{\hat{\alpha}} + \widehat{D}_{\hat{\alpha}}\Lambda'_{\alpha}
\end{align}
where
\begin{align}
D_{\alpha} = \partial_{\alpha} - (\gamma^{a}\theta)_{\alpha}\partial_{a},\quad
\widehat{D}_{\hat{\alpha}} = \partial_{\hat{\alpha}} - (\gamma^{a}\widehat{\theta})_{\hat{\alpha}}\partial_{a}
\end{align}
are the supercovariant derivatives of type IIB superspace.
Other superfields of higher dimensions can be constructed from \(A_{\alpha\hat{\alpha}}\)
and \((D_{\alpha},\widehat{D}_{\hat{\alpha}})\).

To construct a non-linear action whose linearization gives the vertex operator of~(\ref{eqn:IIBvertex}),
one covariantizes as usual \(S_{\text{flat}}+V\) with respect the target space reparameterization
by introducing the supervielbein \(E_{M}^{A}\) (\(M=(m,\mu,\hat{\mu})\), \(A=(a,\alpha,\hat{\alpha})\))
and the curved spacetime coordinate \(Z^{M}=(x^{m},\theta^{\mu},\widehat{\theta}^{\hat{\mu}})\):
\begin{align}
\label{eqn:IIBaction}
S &= {1\over\pi\alpha'}\int\mathd^{2}z
\Bigl(
{1\over2}(G_{MN} + B_{MN})\partial Z^{M}\widebar{\partial} Z^{N} \nonumber\\
&\qquad{} + d_{\alpha} \widebar{\partial} Z^{M} E_{M}^{\alpha}
{} + \partial Z^{M} \widehat{d}_{\hat{\alpha}} E_{M}^{\hat{\alpha}}
{} + d_{\alpha}\widehat{d}_{\hat{\beta}}P^{\alpha\hat{\beta}} \nonumber\\
&\qquad{} + (w_{\alpha}\widebar{\partial}\lambda^{\alpha} + {1\over2}\widebar{\partial} Z^{M}N_{ab} \Omega_{M}^{ab})
{} + (\widehat{w}_{\hat{\alpha}}\partial\widehat{\lambda}^{\hat{\alpha}} + {1\over2}\partial Z^{M} \widehat{N}_{ab} \widehat{\Omega}_{M}^{ab}) \nonumber\\
&\qquad{} + d_{\alpha}\widehat{N}_{ab} C^{\alpha,ab}
{} + \widehat{d}_{\hat{\beta}} N_{ab}\widehat{C}^{\hat{\beta},ab}
{} + {1\over4}N^{ab}\widehat{N}^{cd}R_{abcd}
\Bigr)
\mathperiod
\end{align}
First line is just the standard non-linear sigma model of the Green-Schwarz formalism
in a conformal gauge, where the term with \(G_{MN} = \eta_{ab}E^{a}_{M}E^{b}_{N}\) is the kinetic term
and the one with \(B_{MN}\) is the Wess-Zumino term
(possibly with an integration over an extra dimension).
It is useful to remember that
\(P^{\alpha\hat{\alpha}}\) is a superfield whose lowest component
is the Ramond-Ramond fieldstrength, and \(R_{abcd}\) is a superfield whose lowest component
is the spacetime curvature.

The BRST operator is still given by the expression of the form~(\ref{eqn:flatQB}),
but its action on fields is defined via commutation relations
between \((Z^{M},\lambda^{\alpha},\widehat{\lambda}^{\alpha})\) and their canonical conjugates.
Conditions for this definition to make sense, namely
the conservation of the BRST currents \(\widebar{\partial}(\lambda^{\alpha}d_{\alpha})=\partial(\widehat{\lambda}^{\hat{\alpha}}\widehat{d}_{\hat{\alpha}})=0\)
and nilpotency of the BRST charge,
actually imply supergravity equations of motion for the background superfields~\cite{Berkovits:2001ue}.
Since requiring the kappa symmetry in a generic supergravity
puts the background superfields on-shell
in the Green-Schwarz formalism~\cite{Grisaru:1985fv}\cite{Witten:1985nt},
this is consistent with the expectation that
the kappa symmetry is replaced by the BRST symmetry in the pure spinor formalism.

Also, note that the action of~(\ref{eqn:IIBaction})
can be checked to be BRST invariant if the first line (the ``Green-Schwarz part'')
is assumed to be kappa symmetric~\cite{Oda:2001zm}.
This is not entirely obvious
and means that a Green-Schwarz action in any supergravity background
can be consistently extended to a pure spinor action of the form~(\ref{eqn:IIBaction}).
This observation, on the other hand, does not explain
the equivalence of the two formalisms even at a classical level.

When the Ramond-Ramond superfield \(P^{\alpha\hat{\alpha}}\) is invertible as a \(16\times16\) matrix,
\((d_{\alpha},\widehat{d}_{\hat{\beta}})\) becomes auxiliary and the action~(\ref{eqn:IIBaction})
can be simplified to
\begin{align}
\label{eqn:IIBaction2}
S &= {1\over\pi\alpha'}\int\mathd^{2}z
\Bigl(
{1\over2}(G_{MN} + B_{MN})\partial Z^{M}\widebar{\partial} Z^{N} \nonumber\\
&\qquad{} + (w_{\alpha}\widebar{\nabla} \lambda^{\alpha} + {1\over2}\widebar{\partial} Z^{M}N_{ab} \Omega_{M}^{ab})
{} + (\widehat{w}_{\hat{\alpha}}\partial\widehat{\lambda}^{\hat{\alpha}} + {1\over2}\partial Z^{M} \widehat{N}_{ab} \widehat{\Omega}_{M}^{ab})
{} + {1\over4}N^{ab}\widehat{N}^{cd}R_{abcd}
\Bigr)
\end{align}
for some shifted background superfields.
The action~(\ref{eqn:IIBaction2}) still has a BRST symmetry
and the corresponding charge reads
\begin{align}
\QB &= \int\mathd z \lambda^{\alpha}\partial Z_{M} E_{\alpha}^{M} + \int\mathd\widebar{z}\widehat{\lambda}^{\hat{\alpha}}\widebar{\partial}Z_{M} E_{\hat{\alpha}}^{M}
\mathperiod
\end{align}
It is this form of the action that we shall be using in
our analysis of strings in an $AdS_{5}\times S^{5}$ background,
since the Ramond-Ramond flux is non-degenerate (and constant)
in the background.

\subsection{\texorpdfstring{$AdS_{5}\times S^{5}$}{AdS5 x S5} background}
\label{subsec:PSaction}

For a maximally supersymmetric $AdS_{5}\times S^{5}$ background with constant Ramond-Ramond flux,
one may use the Metsaev-Tseytlin construction~\cite{Metsaev:1998it}
to explicitly write down the background superfields
in the action of~(\ref{eqn:IIBaction2})~\cite{Berkovits:2000fe}.
A reason why it works is that an appropriate superspace can be
written as a supercoset of the form \(G/H=PSU(2,2|4)/(SO(4,1)\times SO(5))\).

\subsubsection{Metsaev-Tseytlin coset construction of Green-Schwarz action for \texorpdfstring{$AdS_{5}\times S^{5}$}{AdS5 x S5}}
\label{subsubsec:MT}

The basic building block for the Metsaev-Tseytlin coset construction is
the left invariant Maurer-Cartan $1$-form
\(\widetilde{J}=\widetilde{g}^{-1}\mathd \widetilde{g}\) (\(\widetilde{g}\in G\)) on \(G\),
or more precisely its pull-back to \(G/H\) via a section \(g\colon G/H\to G\):
\begin{align}
J &= g^{-1}\mathd g
\mathperiod
\end{align}
To construct an action on the coset \(G/H\) using \(J\),
an \(H\) gauge invariance shall be introduced
to make the choice of the section \(g\) irrelevant.
For an application to the \(AdS_{5}\times S^{5}\) superstring relevant groups are
\(G=PSU(2,2|4)\) and \(H=SO(4,1)\times SO(5)\)
and \(J\) takes values in the Lie algebra $\mathfrak{g}=\mathfrak{psu}(2,2|4)$.\footnote{%
See appendix~\ref{sec:appendix} for our conventions for $\mathfrak{psu}(2,2|4)$.}

If one regards $g=g(\tau,\sigma)$ as a function on a worldsheet
with values in the section \(G/H\subset G\),
the $1$-form \(J\) becomes a current on the worldsheet.
The Maurer-Cartan equation can then be pulled back to the worldsheet
and it implies that \(J\) satisfies
\begin{align}
\partial_{+}J_{-}-\partial_{-}J_{+}+[J_{+},J_{-}]=0
\end{align}
where \(J_{\pm}={1\over2}(J_{\tau}\pm J_{\sigma})\) are lightcone components of the current \(J\).

The current \(J\) carries a local \(H\) action and a global \(G\) action
that are inherited from the section \(g\colon G/H \to G\).
Namely, under a local \(H\) transformation of \(g\) defined by
\begin{align}
\label{eqn:LocalH}
g \to g h(\tau,\sigma),\quad h=h(\tau,\sigma) \in H
\end{align}
\(J\) transforms as
\begin{align}
J \to h^{-1}\mathd h + h^{-1}Jh
\end{align}
and under a global \(G\) transformation of \(g\) defined by
\begin{align}
\label{eqn:GlobalG}
g(x) \to g(xa) = ag(x)h(a;\tau,\sigma)^{-1},\quad x\in G/H,\; a\in G,\; h(a;\tau,\sigma)\in H
\end{align}
\(J\) transforms as
\begin{align}
J \to hJh^{-1} - (\mathd h)h^{-1}
\mathperiod
\end{align}
So, \(J\) is invariant under the global \(G\) transformation
up to a compensating \(H\) gauge transformation.

For the case at hand, Lie algebra \(\mathfrak{g}\) of \(G=PSU(2,2|4)\)
admits a \(\mathbb{Z}_{4}\) grading,
\begin{align}
\mathfrak{g} = \bigoplus_{i=0}^{3}\mathfrak{g}^{i},\quad
[\mathfrak{g}^{i},\mathfrak{g}^{j}] \subset \mathfrak{g}^{i+j},\quad
i,j\in\mathbb{Z}_{4}
\end{align}
and the degree zero piece \(\mathfrak{g}^{0}\)
is nothing but the Lie algebra of the denominator \(H=SO(4,1)\times SO(5)\).
Hence, if one decomposes the Metsaev-Tseytlin current by
the \(\mathbb{Z}_{4}\) grading as
\begin{align}
\begin{split}
J &= J^{A}T_{A} = J^{0} + J^{1} + J^{2} + J^{3},\quad J^{i} \in \mathfrak{g}^{i} \mathcomma\\
J^{0} &= J^{ab}L_{ab},\quad
J^{1} = J^{a}Q_{\alpha},\quad
J^{2} = J^{a}P_{a},\quad
J^{3} = J^{\hat{\alpha}}Q_{\hat{\alpha}}
\end{split}
\end{align}
the local \(H\) transformations can be refined as
\begin{align}
\label{eqn:currentGauge1}
J^{0} \to h^{-1}\mathd h + h^{-1}J^{0}h\quad\text{and}\quad
J^{i} \to h^{-1}J^{i}h\quad (i=1,2,3)
\mathperiod
\end{align}
This refinement facilitates the construction of a \(G\)-invariant action
on a supercoset \(G/H\), just like in the case of a symmetric coset space.

Since the currents \(J^{i}\) (\(i=1,2,3\)) transforms homogeneously
under the \(H\) gauge transformation of~(\ref{eqn:LocalH})
an action of the form
\begin{align}
\label{eqn:ginvAction}
\int\mathd^{2}\sigma\str\Bigl({1\over2}J^{2}_{+}J^{2}_{-} + aJ^{1}_{+}J^{3}_{-} + bJ^{3}_{+}J^{1}_{-}\Bigr)
\end{align}
for any constants \(a,b\) is invariant under
the global \(G\) action of~(\ref{eqn:GlobalG})
and the local \(H\) action of~(\ref{eqn:LocalH}).
However, the coset \(G/H\) has \(32\) (too many) fermionic dimensions
and one does not expect~(\ref{eqn:ginvAction}) to describe a superstring
except perhaps at some special values of \((a,b)\).
Just as in a flat superspace,
to construct a superstring model
using the coset action of~(\ref{eqn:ginvAction}),
one has to kill a half of fermionic coordinates
either by introducing a fermionic local symmetry (kappa symmetry)~\cite{Green:1983wt},
or by coupling it to  appropriate bosonic ghosts (like pure spinors)~\cite{Berkovits:2000fe}.
Remarkably, both can be done.

In the works of Metsaev and Tseytlin~\cite{Metsaev:1998it}
and Berkovits et al.~\cite{Berkovits:1999zq},
it was found that a kappa symmetric Green-Schwarz action
in a conformal gauge
can indeed be written in the form~(\ref{eqn:ginvAction})
and is essentially unique (\(a=-b=\pm1/4\)):
\begin{align}
\label{eqn:GSAdSaction}
S_{\text{GS}} &= {R^{2}\over\pi\alpha'}\int\mathd^{2}\sigma
{} \str\Bigl({1\over2}J^{2}_{+}J^{2}_{-}
{} - {1\over4}(J^{1}_{+}J^{3}_{-} - J^{3}_{+}J^{1}_{-})
\Bigr)
\mathperiod
\end{align}
That the Wess-Zumino term can be written as an integration
over the two dimensional worldsheet follows from the fact
that \(\mathfrak{psu}(2,2|4)\) admits
a \(\mathbb{Z}_{4}\) automorphism~\cite{Berkovits:1999zq}.
The ``radius'' parameter \(R\) is related to the number \(N\) of D3-branes that
source the Ramond-Ramond flux supporting \(AdS_{5}\times S^{5}\),
but the integrality of \(N\) cannot be probed by an elementary string.
From now on we set the radius \(R\) in the unit of \(\sqrt{\alpha'}\) to be one.
In the context of the AdS/CFT correspondence,
the semiclassical parameter \(\alpha'\) then is related to the 't~Hooft coupling \(\lambda\)
of the \(\calN=4\) super Yang-Mills theory as \(\alpha' \sim 1/\sqrt{\lambda}\).

Since the Green-Schwarz action of~(\ref{eqn:GSAdSaction}) is written in a conformal gauge,
it is understood to be accompanied by Virasoro constraints
\begin{align}
\label{eqn:GSVirNaive}
T = {1\over2\alpha'}\str(J_{+}^{2}J_{+}^{2}) \approx 0,\quad \widebar{T} = {1\over2\alpha'}\str(J_{-}^{2}J_{-}^{2}) \approx 0
\mathperiod
\end{align}
Note that the second term of~(\ref{eqn:GSAdSaction})
is a topological Wess-Zumino term
(i.e.\ does not couple to worldsheet metric)
and hence does not contribute to the stress tensors.
However, the Green-Schwarz action have more constraints than
the Virasoro constraints of~(\ref{eqn:GSVirNaive})
and separation of the first and second class constraints
makes it more natural to improve the naive Virasoro constraints
so that they become first class.
The improved Virasoro constraints are then closely related to
the stress tensor of the pure spinor formalism.

\subsubsection{Pure spinor action for \texorpdfstring{$AdS_{5}\times S^{5}$}{AdS5 x S5}}

In subsection~\ref{subsec:sugraBG} we explained a relation between
Green-Schwarz action and pure spinor action in an arbitrary
supergravity background.
One can find a pure spinor action in an $AdS_{5}\times S^{5}$ background
by applying the argument to the Metsaev-Tseytlin action.
In the ``second order'' form it reads\footnote{%
Here, we have judiciously used the opposite sign for
the Wess-Zumino term in \(S_{\text{GS}}\) with respect to the one given in~(\ref{eqn:GSAdSaction})
because it is the variables in~(\ref{eqn:PSAdSaction1}) that have a simple relation
to the Green-Schwarz variables of~(\ref{eqn:GSAdSaction}).
Otherwise the relation between the variables of the two formalisms
gets twisted by an automorphism \(\mathfrak{g}^{1}\leftrightarrow\mathfrak{g}^{3}\)
of \(\mathfrak{psu}(2,2|4)\).
Of course, this is a matter of convention but we find it prettier this way.}
\begin{align}
\label{eqn:PSAdSaction0}
S &= S_{\text{GS}} + {1\over\pi\alpha'}\int\mathd^{2}\sigma\str\Bigl(J_{+}^{3}J_{-}^{1}\Bigr)
{} + S_{\text{gh}} \\
\label{eqn:PSAdSaction1}
&= {1\over\pi\alpha'}\int\mathd^{2}\sigma\str\Bigl({1\over2}J^{2}_{+}J^{2}_{-} + {1\over4}J_{+}^{1}J_{-}^{3} + {3\over4}J_{+}^{3}J_{-}^{1}\Bigr)
+ S_{\text{gh}}
\mathperiod
\end{align}
where
\begin{align}
\label{eqn:PSAdSactionghost}
\begin{split}
S_{\text{gh}}
&= {1\over\pi\alpha'}\int\mathd^{2}\sigma\str\Bigl( w^{3}[D_{-},\lambda^{1}] + \widehat{w}^{1}[D_{+},\widehat{\lambda}^{3}] - N\widehat{N}\Bigr)
\mathcomma\\
(D_{\pm} &= [\partial_{\pm} + J^{0}_{\pm},\;\cdot\;] )
\end{split}
\end{align}
describes the contribution of pure spinor ghosts and their coupling to the ``matter'' sector.
The second term in~(\ref{eqn:PSAdSaction0}) comes from integrating out
the auxiliary fields \((d_{\alpha},\widehat{d}_{\hat{\alpha}})\)
as explained at the end of subsection~\ref{subsec:sugraBG}.
In the ``ghost'' action \(S_{\text{gh}}\)~(\ref{eqn:PSAdSactionghost})
we have introduced pure spinor variables as supermatrices
\begin{align}
\lambda^{1} &= \lambda^{\alpha}T_{\alpha} \in \mathfrak{g}^{1},\quad \widehat{\lambda}^{3} = \widehat{\lambda}^{\hat{\alpha}}T_{\hat{\alpha}} \in \mathfrak{g}^{3}
\end{align}
satisfying \(SO(4,1)\times SO(5)\) pure spinor constraint
\begin{align}
\{\lambda^{1},\lambda^{1}\} = \lambda^{\alpha}\gamma^{a}_{\alpha\beta}\lambda^{\beta} = 0,\quad
\{\widehat{\lambda}^{3},\widehat{\lambda}^{3}\} = \widehat{\lambda}^{\hat{\alpha}}\gamma^{a}_{\hat{\alpha}\hat{\beta}}\widehat{\lambda}^{\hat{\beta}} = 0
\mathperiod
\end{align}
Since the pure spinor ghosts are bosonic, supermatrices \(\lambda^{1}\) and \(\widehat{\lambda}^{3}\)
have a wrong Grassmann parity.
We have also introduced the conjugates to \(\lambda^{1}\) and \(\widehat{\lambda}^{3}\)
\begin{align}
w^{3} &= \eta^{\alpha\hat{\alpha}}w_{\alpha}T_{\hat{\alpha}} \in \mathfrak{g}^{3},\quad
\widehat{w}^{1} = \eta^{\hat{\alpha}\alpha}\widehat{w}_{\hat{\alpha}}{T}_{{\alpha}} \in \mathfrak{g}^{1}
\end{align}
and Lorentz (\(SO(4,1)\times SO(5)\)) generators of the pure spinor sector
\begin{align}
N &= -\{w^{3},\lambda^{1}\},\quad \widehat{N} = -\{\widehat{w}^{1},\widehat{\lambda}^{3}\}
\mathperiod
\end{align}

Note that the matter sector of pure spinor superstring action~(\ref{eqn:PSAdSaction1}) is not kappa symmetric
since Green-Schwarz action of~(\ref{eqn:GSAdSaction}) is the unique such action.
Another important difference
is that the pure spinor action is not accompanied by Virasoro constraints
even though it is written in a ``conformal gauge''.
In pure spinor formalism, both the kappa symmetry and the Virasoro constraint
are replaced by a BRST symmetry.

\subsection{\texorpdfstring{$PSU(2,2|4)$}{PSU(2,2|4)} symmetry and Noether current}

The local \(H=SO(4,1)\times SO(5)\) transformation of~(\ref{eqn:LocalH})
and the global \(G=PSU(2,2|4)\) transformation of~(\ref{eqn:GlobalG})
can be extended to the pure spinor sector in a way that the action is invariant.
The coupling of pure spinors to the connection \(J^{0}\) implies
that the former is
\begin{align}
g &\to g h(\tau,\sigma),\quad
(w,\lambda,\widehat{w},\widehat{\lambda}) \to h(\tau,\sigma)^{-1}(w,\lambda,\widehat{w},\widehat{\lambda})h(\tau,\sigma),\quad h(\tau,\sigma)\in H
\end{align}
and the latter is
\begin{align}
g &\to agh(a;\tau,\sigma)^{-1},\quad
(w,\lambda,\widehat{w},\widehat{\lambda}) \to h(a;\tau,\sigma)(w,\lambda,\widehat{w},\widehat{\lambda})h(a;\tau,\sigma)^{-1},\quad a \in G,\; h(a;\tau,\sigma)\in H
\mathperiod
\end{align}

The Noether current associated with the \(PSU(2,2|4)\) symmetry
can be computed in a standard manner,
and is given by
\begin{align}
\label{eqn:NoetherCurrent}
\begin{split}
j &= (j_{+},j_{-}) = j^{A}T_{A} \in \mathfrak{psu}(2,2|4)
\mathcomma\\
j_{+} &= g(J^{2}_{+} + {1\over2}J^{1}_{+} +{3\over2}J^{3}_{+} + 2N)g^{-1},\quad
j_{-} = g(J^{2}_{-} + {3\over2}J^{1}_{-}
{} + {1\over2}J^{3}_{-} + 2\widehat{N})g^{-1}
\mathperiod
\end{split}
\end{align}
The normalization of \(j\) here is such that the corresponding conserved charge
is given by \({1\over4\pi\alpha'}\int\mathd\sigma j^{A}_{\tau}\).
Individual components for each \(\mathfrak{psu}(2,2|4)\) generator can
be extracted as
\begin{align}
j^{A} = \eta^{AB}\str(T_{B}\,j)
\end{align}
where \(\eta^{AB}\) is the inverse of the trace metric \(\eta_{AB}=\str(T_{A}T_{B})\).
Of particular importance for us is
the components for \(T_{0},T_{9}\in \mathfrak{g}^{2}\).
Conserved charges associated with them are
the \(AdS\) energy and an angular momentum in \(S^{5}\)
\begin{align}
E = {1\over4\pi\alpha'}\int\mathd \sigma j^{0}_{\tau},\quad
J = {1\over4\pi\alpha'}\int\mathd \sigma j^{9}_{\tau}
\mathperiod
\end{align}

\subsection{BRST symmetry, composite \texorpdfstring{$b$}{b}-ghost and stress tensor}

The pure spinor action of~(\ref{eqn:PSAdSaction1})
is invariant under an on-shell BRST transformation
defined by\footnote{The BRST symmetry can be promoted to an off-shell
symmetry by adding some auxiliary fields~\cite{Berkovits:2007rj,Boussard}.}
\begin{align}
\delta_{\text{B}} g &= g (\lambda^{1}+\widehat{\lambda}^{3}),\quad
\delta_{\text{B}}w^{3} = -J^{3}_{+},\quad \delta_{\text{B}}\widehat{w}^{1} = -J^{1}_{-},\quad
\delta_{\text{B}}\lambda^{1} = \delta_{\text{B}}\widehat{\lambda}^{3} = 0
\mathperiod
\end{align}
On Metsaev-Tseytlin currents, it acts as
\begin{align}
\delta_{\text{B}}J^{0} &= [J^{3},\lambda^{1}] + [J^{1},\widehat{\lambda}^{3}],\quad
\delta_{\text{B}}J^{1} = [D,\lambda^{1}] + [J^{2},\widehat{\lambda}^{3}],
\\
\delta_{\text{B}}J^{2} &= [J^{1},\lambda^{1}] + [J^{3},\widehat{\lambda}^{3}],\quad
\delta_{\text{B}}J^{3} = [D,\widehat{\lambda}^{3}] + [J^{2},\lambda^{1}]
\mathperiod
\end{align}
Associated BRST charge can be written as
a sum of left-moving and right-moving components
\begin{align}
\label{eqn:AdSBRSTcharge}
\QB &= Q + \widebar{Q},\quad
Q = \int\mathd \sigma^{+} \str(\lambda^{1} J^{3}_{+}),\quad \widebar{Q} = \int\mathd \sigma^{-} \str(\widehat{\lambda}^{3}J^{1}_{-})
\end{align}
where \(\partial_{-}\str(\lambda^{1} J^{3}_{+})= \partial_{+}\str(\widehat{\lambda}^{3}J^{1}_{-}) = 0\)
because of the equations of motion.

In any BRST formulation of string theory,
it is crucial to have \(b\) ghost fields that
make stress tensors BRST trivial as in \(\{\QB,b\}=T\), \(\{\QB,\widehat{b}\}=\widebar{T}\).
Since the stress tensors
\begin{align}
\label{eqn:PSAdSstress}
T &= {1\over \alpha'} \str ({1\over2}J^{2}_{+}J^{2}_{+} + J^{1}_{+}J^{3}_{+} + w^{3}[D_{+},\lambda^{1}]),\quad
\widebar{T} = {1\over \alpha'} \str ({1\over2}J^{2}_{-}J^{2}_{-} + J^{1}_{-}J^{3}_{-} + \widehat{w}^{1}[D_{-},\widehat{\lambda}^{3}])
\end{align}
carry ghost number \((0,0)\)
while \(Q\) and \(\widebar{Q}\) carry ghost numbers \((1,0)\) and \((0,1)\),
one needs operators of negative ghost numbers to construct the \(b\) ghosts.
In an \(AdS_{5}\times S^{5}\) background
\((\lambda\widehat{\lambda})\equiv \str(\lambda^{1}\widehat{\lambda}^{3})\) is in the cohomology of \(\QB\),
and it has been argued that it is consistent to allow inverse powers of \((\lambda\widehat{\lambda})\)~\cite{Berkovits:2008ga}.
One can utilize this observation to construct composite \(b\) ghosts with negative
ghost numbers \((-1,0)\) and \((0,-1)\) as~\cite{Berkovits:2008ga,Berkovits:2010zz}
\begin{align}
\begin{split}
b &= {1\over \alpha'}\str\Bigl({ \widehat{\lambda}^{3}[J^{2}_{+},J^{3}_{+}] \over (\lambda\widehat{\lambda}) }
{} - w^{3}J^{1}_{+}
{} + { \{w^{3},\widehat{\lambda}^{3}\}[\lambda^{1},J^{1}_{+}] \over (\lambda\widehat{\lambda}) }\Bigr),
\\
\widehat{b} &= {1\over \alpha'}\str\Bigl({ \lambda^{1}[J^{2}_{-},J^{1}_{-}] \over (\lambda\widehat{\lambda}) }
{} - \widehat{w}^{1}J^{3}_{-}
{} + { \{\widehat{w}^{1},\lambda^{1}\}[\widehat{\lambda}^{3},J^{3}_{-}] \over (\lambda\widehat{\lambda}) }\Bigr)
\end{split}
\end{align}
and it can be checked that these satisfy
\begin{align}
\{Q,b\} = T,\quad \{\widebar{Q},\widehat{b}\} = \widebar{T},\quad
\{Q,\widehat{b}\} = \{\widebar{Q},b\} = 0
\mathperiod
\end{align}
Note that \(b\) and \(\widehat{b}\) are actually invariant under
\(\delta_{\Omega} w^{3} = \{\Omega^{2},\lambda^{1}\}\) and
\(\delta_{\Omega} \widehat{w}^{1} = \{\Omega^{2},\widehat{\lambda}^{3}\}\)
for an arbitrary operator \(\Omega^{2}\)
and that, although \(b\) is not purely left-moving
and \(\widehat{b}\) is not purely right-moving,
\(\partial_{-}b\) and \(\partial_{+}\widehat{b}\) are BRST trivial~\cite{Berkovits:2010zz}.

A remark is in order.
The action of~(\ref{eqn:PSAdSaction1}) can be naively coupled to worldsheet gravity
and the stress tensor of~(\ref{eqn:PSAdSstress}) are the ones
that one would obtain from this coupling.
However, as mentioned earlier,
the action of~(\ref{eqn:PSAdSaction1}) should not be regarded as arising from
gauge fixing this naive reparameterization invariant action,
for that would imply that the stress tensor is a constraint.
If one wishes to start from a reparameterization invariant action,
the correct starting point should rather be the classical Green-Schwarz action.
Studies along this line in a flat background
tell us that the pure spinor variables arise as bosonic ghosts
for the kappa symmetry, and that one should think of the fundamental $bc$-ghosts
to be ``integrated out'' from the theory,
effectively getting replaced by one of the pure spinor constraints~\cite{Berkovits:2000nn,Aisaka:2005vn}.

\subsection{Classical equations of motion}
\label{subsec:classEOM}

Equations of motion for both Green-Schwarz and pure spinor superstrings
can be readily computed from their actions~(\ref{eqn:GSAdSaction}) and (\ref{eqn:PSAdSaction1}).

\paragraph{Green-Schwarz}

Classical equations of motion for the Green-Schwarz superstring
in an \(AdS_{5}\times S^{5}\) background is well known.
In a conformal gauge they read
\begin{alignat}{2}
\label{eqn:GSclassEOM}
[D_{-},J_{+}^{2}] + [J_{-}^{1},J_{+}^{1}] &= 0,&\qquad
[D_{+},J_{-}^{2}] + [J_{+}^{3},J_{-}^{3}] &= 0
\mathcomma
\\
[J_{-}^{2},J_{+}^{3}]&=0,&\qquad
[J_{+}^{2},J_{-}^{1}]&= 0
\end{alignat}
where, as before, the spin covariant derivatives are defined as
$D_{\pm} = \partial_{\pm} + [J_{\pm}^{0},\;\cdot\;]$.
These are understood to be supplemented by
the Maurer-Cartan equations
\begin{align}
\partial_{+} J^{i}_{-} - \partial_{-} J^{i}_{+}
{} + \sum_{j+k=i}[J^{j}_{+},J^{k}_{-}] = 0,\quad (i\in\mathbb{Z}_{4})
\end{align}
and by the Virasoro constraint
coming from a choice of the conformal gauge
\begin{align}
\str(J^{2}_{+}J^{2}_{+}) = \str(J^{2}_{-}J^{2}_{-}) = 0
\mathperiod
\end{align}

\paragraph{Pure spinor}

The currents from the matter sector of the pure spinor formalism
satisfy the same set of Maurer-Cartan equations
as the ones in the Green-Schwarz formalism,
but their equations of motion are different:
\begin{alignat}{2}
\label{eqn:PSclassEOM}
[D_{-}-\widehat{N},J_{+}^{2}] + [J_{-}^{1},J_{+}^{1}]  &=[J^{2}_{-},N] ,&\qquad
[D_{+}-N,J_{-}^{2}] + [J_{+}^{3},J_{-}^{3}] &= [J^{2}_{+},\widehat{N}]
\mathcomma
\\
[D_{-}-\widehat{N},J^{3}_{+}]  &= [J^{3}_{-},N],&\qquad
[D_{+}-N,J_{-}^{1}]  &= [J^{1}_{+},\widehat{N}]
\mathperiod
\end{alignat}
If one ignores ghost contributions,
the equations of motion for the bosonic current \(J^{2}_{\pm}\)
reduce to that of the Green-Schwarz formalism.
On the other hand,
the equations of motion for the fermionic currents \(J^{1}_{\pm}\) and \(J^{3}_{\pm}\)
take the forms of covariant constancy conditions
even after dropping the ghost contributions
and do not reduce to the ``algebraic'' equations of motions
of the Green-Schwarz formalism.

Equations of motion for the pure spinor ghost variables are
\begin{alignat}{2}
[D_{-} - \widehat{N} ,\lambda^{1}] &= 0,&\qquad [D_{+} - N, \widehat{\lambda}^{3}] &= 0
\mathcomma\\
[D_{-} - \widehat{N}, w^{3}] &= 0,&  [D_{+} - N, \widehat{w}^{1}] &= 0
\mathperiod
\end{alignat}
The equations for \((w^{3},\widehat{w}^{1})\)
can be replaced by that for the gauge invariant Lorentz currents
\begin{align}
[D_{-} - \widehat{N}, N] &= 0,\qquad [D_{+} - N, \widehat{N}] = 0
\mathperiod
\end{align}

Unlike in the Green-Schwarz formalism,
the Virasoro condition is not a part of the equations of motion.
Nevertheless, in a semiclassical setup,
it is still true that the ``classical solution''
around which one studies small fluctuations
should have vanishing worldsheet energy and momentum
(\(L_{0}\pm\widebar{L}_{0}\)),
since the Virasoro currents \(T\) and \(\widebar{T}\) are BRST exact.

\section{Semiclassical pure spinor superstring in \texorpdfstring{$AdS_{5}\times S^{5}$}{AdS5 x S5} background}
\label{sec:semiclassicalPS}

We now turn to the main topic of the present article.
Our primary goal is to explain the reason why
the one-loop correction to classical string energy computed
using the pure spinor formalism agrees with that from the Green-Schwarz formalism.
For simplicity, we shall restrict ourselves to a simple family of
classical solutions (defined in section~\ref{subsec:family}),
but we believe that the pattern that connects the two formalisms
stay the same for a broader class of solutions.

The structure of our argument is as follows.
After developing some semiclassical formulas for the pure spinor superstring
around a generic classical solution,
we show that, for a certain class of solutions,
the one-loop correction to spacetime energy
comes entirely from the zero-point ``energy'' of worldsheet fluctuations.
The zero-point ``energy'' is the normal ordering constant in the Hamiltonian
of quadratic fluctuations, and can be computed from the one-loop partition function
on the worldsheet.
To argue that the one-loop partition functions of Green-Schwarz and pure spinor
formalisms agree, we analyze the equations of motion for fluctuations of
the latter and identify Green-Schwarz like degrees of freedom.
Morally speaking, those degrees of freedom are
related to the BRST cohomology of fluctuations and yield the same zero-point ``energy''
as the Green-Schwarz fluctuations.
The remaining degrees of freedom, which are decoupled from the Green-Schwarz like ones,
have a trivial partition function and do not contribute to the zero-point ``energy''.

\subsection{Comparison of semiclassical analyses for Green-Schwarz and pure spinor formalisms}

As we have reviewed in the previous section,
compared to the Green-Schwarz formalism, the pure spinor formalism
has an extended set of fields and the Virasoro and kappa symmetries
are replaced by a BRST symmetry.
To compare semiclassical analyses in Green-Schwarz and pure spinor formalisms,
one has to identify classical solutions of both sides
and compare the structure of small fluctuations around them.

From the forms of classical equations of motion~(subsection~\ref{subsec:classEOM}),
one finds that a purely bosonic solution of the Green-Schwarz formalism is
automatically a solution of the pure spinor formalism
(with a trivial ghost profile).
However, it is not clear if all classical solutions
of the pure spinor formalism can be obtained in this way.
In this article, we shall leave the complete comparison of the space
of classical solutions
along the line of~\cite{Beisert:2005bm}
as an interesting open question.

So in the discussion that follows,
we pick a solution of the Green-Schwarz formalism and
regard it as the solution of the pure spinor formalism
describing the same classical string.

Since the Green-Schwarz action in a conformal gauge comes
with Virasoro and kappa symmetries, fluctuations around a classical solution
have to respect certain constraints.
The presence of the kappa symmetry manifests itself in the semiclassical analysis
as a degeneracy of fermionic propagators. Namely, one half of the fermionic fluctuations
does not propagate and one may simply freeze these fluctuations to
deal with the kappa symmetry.
The Virasoro constraint implies that two of ten bosonic fluctuations are
functionals of others, and normally the two fluctuations are removed
by either imposing a lightcone gauge or a static gauge condition.

After properly dealing with the constraints,
one may in principle quantize the quadratic fluctuations and compute semiclassical quantities.
The classical solution is identified with
the ground state \(|\Omega\rangle\)
of the worldsheet Hamiltonian \(H_{2}\) for the quadratic fluctuations,
and a semiclassical correction to the spacetime energy of the solution
can be computed as
\begin{align}
\Delta E({\Omega}) = \langle \Omega| (E-\underline{E})|\Omega\rangle
\mathperiod
\end{align}
Here, \(E\) on the right hand side is the Noether charge for the \(AdS\) time translation
written in terms of fluctuations and \(\underline{E}\) denotes
its classical value.
For the class of solutions defined in section~\ref{subsec:family},
this quantity can be related to the expectation value of the worldsheet
Hamiltonian \(H_{2}\) by imposing Virasoro constraint on fluctuations~\cite{Frolov:2002av}.
This is a good fortune because
one can bypass the explicit quantization of fluctuations
when computing \(\Delta E(\Omega)\).

As an aside, let us mention that one may ignore the fluctuations of Goldstone modes
to the one-loop approximation and that
a quantum state \(|\Psi\rangle\) with some excitations over \(|\Omega\rangle\)
represents a string state with slightly higher energy.
Quantization of Goldstone modes is interesting
(this should turn the ground state to a multiplet of spontaneously broken
global symmetries),
and is certainly important for two-loops and beyond.
We, however, do not inquire into these issues in this article.

\bigskip
In the pure spinor formalism, the procedure for the semiclassical analysis is similar
but now the Virasoro and kappa symmetries are replaced by a BRST symmetry.

When performing a semiclassical analysis for a BRST system in general,
it is useful to keep the following geometric picture in mind~(cf.~\cite{SchwarzSC}).
Presence of a (on-shell) nilpotent BRST symmetry implies that a critical point of the action
in the space of fields
belongs either to a trivial orbit (BRST singlet)
or a non-trivial orbit with zero volume (BRST doublet).
A ``classical solution'' around which one performs a semiclassical analysis
has to be a solution to the equations of motion
and at the same time a BRST singlet.
When a solution is a BRST singlet,
the BRST symmetry induces a nilpotent action on fluctuations around the solution.
So one gets a new BRST system of fluctuations
and the ground state \(|\Omega\rangle\) and excited states \(|\Psi\rangle\)
are defined as BRST cohomologies.
Semiclassical quantization of fluctuations of a BRST system
around a ``classical solution'' is
conceptually simpler than that of a gauge invariant system because all
the problems with degenerate phase space of the latter
are already taken care of by the BRST symmetry.

Coming back to the relation between Green-Schwarz and pure spinor formalisms,
one expects that a quantum state \(|\Psi\rangle\) of the former
can be mapped to a BRST cohomology class of the latter.
This mapping should allow one to directly compare the one-loop corrections
\(\Delta E({\Psi}) = \langle \Psi|(E-\underline{E})|\Psi\rangle\)
in the two formalisms.
Unfortunately, however, it is not necessarily easy to show the equivalence in this way,
just because quantization of fluctuations around a given classical solution
could be too hard.
In general, both kinetic and mass terms are not constant
and moreover have complicated mixing,
so quantization is not easy even for the lightcone Green-Schwarz formalism.

But if one is mainly interested in comparing one-loop corrections \(\Delta E(\Omega)\)
to the energies of the classical solution,
explicit quantization can be sometimes circumvented.
As mentioned above, there is a family of classical solutions
for which one-loop energy corrections
are related to expectation values of their worldsheet Hamiltonians \(H_{2}\),
both in Green-Schwarz and pure spinor formalisms.
Then, the equivalence of the two formalisms (as far as \(\Delta E(\Omega)\) is concerned)
is reduced to a simpler problem of comparing one-loop partition functions.
In subsections~\ref{subsec:fermifluc} and~\ref{subsec:cmp1loop}
we study equations of motions for fluctuations in Green-Schwarz and pure spinor formalisms
and argue that their one-loop partition functions around
the classical solutions of subsection~\ref{subsec:family} do agree.

\subsection{Quadratic fluctuations}

Computations of semiclassical quantities
can be done by using a background field method.
For a sigma model on a group manifold,
a convenient way to separate the worldsheet variable $g(\tau,\sigma)$ to
its background value $\underline{g}(\tau,\sigma)$ and small fluctuations $X(\tau,\sigma)\in\mathfrak{g}$
around it is as
\begin{align}
\label{eqn:BGsplit}
g = \underline{g}\mathe^{X}
\mathperiod
\end{align}
To perform a consistent semiclassical analysis,
\(X\) is understood to be a quantity of order \(\sqrt{\alpha'}\).
When the sigma model is on a coset \(G/H\),
\(\underline{g}\) is a coset representative
and the small fluctuation \(X\) takes values in a subspace of \(\mathfrak{g}\).
Identification~(\ref{eqn:BGsplit}) may require a compensating \(H\) gauge transformation
which, however, is irrelevant for gauge invariant quantities like action.
For the case at hand,
the fluctuation $X$ can be split according to the $\mathbb{Z}_{4}$ grading
of $\mathfrak{g}=\mathfrak{psu}(2,2|4)$
and we choose it to have the components orthogonal to
\(\mathfrak{g}^{0} (= \mathfrak{h})\):
\begin{align}
X = \bigoplus_{i=1}^{3}X^{i},\quad X^{i}\in \mathfrak{g}^{i}
\mathperiod
\end{align}

For simplicity, we assume the background to be purely bosonic and ghost free
(i.e.~no background values for the fermionic currents \((J^{1},J^{3})\)
and the ghosts).

\subsubsection{Quadratic action}

Expansion of the coset action of the form~(\ref{eqn:ginvAction})
to quadratic order in fluctuations is straightforward.
Vast simplification for the end result occur precisely
when the relative coefficients of \(J^{1}_{+}J^{3}_{-}\) and
\(J^{3}_{+}J^{1}_{-}\) with respect to \({1\over2}J^{2}_{+}J^{2}_{-}\)
are either as in the Green-Schwarz action~(\ref{eqn:GSAdSaction})
or as in the pure spinor action~(\ref{eqn:PSAdSaction1}).
Moreover, the fluctuation actions for these two cases
bear a striking resemblance to each other.

\paragraph{Green-Schwarz}

To the quadratic order, there is no mixing of
bosonic and fermionic fluctuations,
so the quadratic action is of the form
\begin{align}
S^{\text{GS}}_{2} &= S^{\text{GS}}_{2B} + S^{\text{GS}}_{2F}
\end{align}
where
\begin{align}
\label{eqn:GSS2B}
S^{\text{GS}}_{2B}
&= {1\over2\pi\alpha'}
\int\mathd^{2}\sigma\str\bigl(
[D_{+},X^{2}][D_{-},X^{2}]
{} - [\underline{J}_{+}^{2},X^{2}][\underline{J}_{-}^{2},X^{2}]
\bigr)
\mathcomma
\\
\label{eqn:GSS2F}
S^{\text{GS}}_{2F}
&= -{1\over2\pi\alpha'}\int\mathd^{2}\sigma \str\bigl([D_{+},X^{1}][\underline{J}_{-}^{2},X^{1}]
{} + [\underline{J}_{+}^{2},X^{3}][D_{-},X^{3}]
{} + 2[\underline{J}_{+}^{2},X^{3}][\underline{J}_{-}^{2},X^{1}]\bigr)
\mathperiod
\end{align}
Here and hereafter, \(\underline{J}_{\pm}\equiv\underline{g}^{-1}\partial_{\pm}\underline{g}\) denotes
the background values of the current \(J_{\pm}\).

A characteristic feature of \(S_{2F}^{\text{GS}}\) is that it has a first order kinetic term.
On a slightly closer inspection one finds that actually
one half of the fermionic fluctuation modes are absent from \(S^{\text{GS}}_{2F}\).
(Roughly speaking, the classical Virasoro constraint implies that
matrices representing \([\underline{J}^{2}_{\pm},\;\cdot\;]\)
have half maximal rank and project out one halves of \(X^{1}\) and \(X^{3}\).)
Of course, this reflects the fact
that the Green-Schwarz action has a kappa symmetry.

\paragraph{Pure spinor}

Since we are assuming that the background values for pure spinor ghosts are trivial,
the quadratic action for the fluctuations is of the form
\begin{align}
S^{\text{PS}}_{2} &= S^{\text{PS}}_{2B} + S^{\text{PS}}_{2F} + S^{\text{PS}}_{2G}
\end{align}
where \(S^{\text{PS}}_{2B}\) is the same as \(S^{\text{GS}}_{2B}\) of Green-Schwarz formalism~(\ref{eqn:GSS2B}) and
\begin{align}
S^{\text{PS}}_{2F}
&= {1\over2\pi\alpha'}\int\mathd^{2}\sigma \str\bigl(
{} 2[D_{+},X^{3}][D_{-},X^{1}]
{} + [\underline{J}_{+}^{2},X^{1}][D_{-},X^{1}]
{} + [D_{+},X^{3}][\underline{J}_{-}^{2},X^{3}]
\bigr)
\mathcomma
\\
S^{\text{PS}}_{2G}
&= {1\over\pi\alpha'}\int\mathd^{2}\sigma \str\bigl(
{} w[D_{-},\lambda] + \widehat{w}[D_{+},\widehat{\lambda}]
{} \bigr)
\mathperiod
\end{align}

Since the fluctuation actions for the bosonic modes $X^{2}$
in Green-Schwarz and pure spinor formalisms are the same,
their contributions to the semiclassical partition functions of
the Green-Schwarz and pure spinor formalisms can be related trivially.
Of course, constraint structures for the fluctuations
are different (Virasoro in Green-Schwarz and BRST in pure spinor),
but it just implies that contributions of unphysical fluctuations along ``lightcone directions''
to physical quantities
get neutralized by different fermionic fluctuations
(reparameterization ghosts in Green-Schwarz and unphysical fermionic
fluctuations in pure spinor).
We therefore focus on more interesting
fermionic fluctuations $(X^{1},X^{3})$ in the following discussions.

Note that the kinetic term for the fermionic fluctuations in \(S^{\text{PS}}_{2F}\)
is of second order and non-degenerate.
This is in sharp contrast to the case of Green-Schwarz.
On the other hand, the appearance of \(S^{\text{PS}}_{2F}\) here is rather similar
to \(S^{\text{GS}}_{2F}\)~(\ref{eqn:GSS2F}) of the Green-Schwarz formalism and can be obtained by formally replacing
the ``mass term'' in \(S^{\text{GS}}_{2F}\) by the second order kinetic term.

\subsubsection{Linearized equations of motion}
\label{subsubsec:lineom}

To compare the structures of fluctuations of Green-Schwarz and
pure spinor formalisms, it is useful to compare their
equations of motions.
We record them here for future use.
We also introduce a component notation
by choosing a basis of \(\mathfrak{g}^{1}\) and \(\mathfrak{g}^{3}\).

\paragraph{Bosonic fluctuations}

Equations of motion for bosonic fluctuation \(X^{2}\in\mathfrak{g}^{2}\)
are the same for Green-Schwarz and pure spinor formalisms:
\begin{align}
[D_{+},[D_{-},X^{2}]]
{} - [\underline{J}^{2}_{+},[\underline{J}^{2}_{-},X^{2}]]
&= 0
\mathperiod
\end{align}
Those modes contribute the same amount to one-loop corrections
in two formalisms and hence are not of primary interest to us.

\paragraph{Green-Schwarz}

By using the classical equations of motion~(\ref{eqn:GSclassEOM}) for the backgrounds
and the Maurer-Cartan equation,
the equations of motion for \(X^{1}\) and \(X^{3}\) are found to be
\begin{align}
[D_{+},[\underline{J}_{-}^{2},X^{1}]]+[\underline{J}^{2}_{-},[\underline{J}^{2}_{+},X^{3}]] &= 0,\quad
[D_{-},[\underline{J}_{+}^{2},X^{3}]]+[\underline{J}_{+}^{2},[\underline{J}^{2}_{-},X^{1}]] = 0
\mathperiod
\end{align}

To study these equations further, it is convenient to take an explicit basis for
\(\mathfrak{g}^{1}\) and \(\mathfrak{g}^{3}\) and denote
\begin{align}
X^{1} &= \theta^{\alpha}T_{\alpha},\quad X^{3} = \widehat{\theta}^{\hat{\alpha}}T_{\hat{\alpha}}
\mathperiod
\end{align}
Actions of \(D_{\pm}\) and \(\underline{J}_{\pm}^{2}\) on
\((\theta^{\alpha},\widehat{\theta}^{\hat{\alpha}})\) can be understood by noting
that the bosonic currents \(\underline{J}^{0}\) and \(\underline{J}^{2}\) are related to
the spacetime spin connection \(\omega_{m}{}^{ab}\) and vielbein \(e_{m}^{a}\)
respectively.
We denote
\begin{align}
(\rho_{\pm})_{\alpha\beta} &\equiv \partial_{\pm}x^{m}e_{m}^{a}(\gamma_{a})_{\alpha\beta},\quad
(\rho_{\pm})^{\alpha\beta} \equiv \partial_{\pm}x^{m}e_{m}^{a}(\gamma_{a})^{\alpha\beta}
\end{align}
where \(\gamma\)'s are \(SO(4,1)\times SO(5)\) gamma matrices.
Spinor indices can be raised and lowered
using the invariant spinor metric \(\eta_{\alpha\hat{\alpha}}=-\eta_{\hat{\alpha}\alpha}\) coupling \(\mathfrak{g}^{1}\) and \(\mathfrak{g}^{3}\)
and its inverse.
We often omit spinor indices assuming that they are contracted appropriately.
It is useful to remember that the classical equations of motion for the background implies
\([D_{\pm},\rho_{\mp}] = 0\)
and that the Virasoro condition implies
\(\rho_{+}\rho_{+} = \rho_{-}\rho_{-} = 0\).
Actually, \(\rho_{\pm}\) have half the maximal ranks
so they act as projectors on spinors.

In terms of $(\theta^{\alpha},\widehat{\theta}^{\hat{\alpha}})$ the equations of motion can be written as
\begin{align}
\label{eqn:GSflucfermi}
D_{+}(\eta \rho_{-}\theta)^{\alpha} - {1\over2}(\eta\rho_{-})^{\alpha}{}_{\hat{\beta}}(\eta\rho_{+}\widehat{\theta})^{\hat{\beta}} &= 0\,,\quad
D_{-}(\eta \rho_{+}\widehat{\theta})^{\hat{\alpha}} + {1\over2}(\eta\rho_{+})^{\hat{\alpha}}{}_{\beta}(\eta\rho_{-}\theta)^{\beta} = 0
\end{align}
where
\(D_{\pm} = \partial_{\pm} - {1\over4}\omega_{\pm}{}^{ab}\gamma_{ab}\)
denotes the action of the covariant derivative \([D_{\pm},\;\cdot\;]\) on spinors.
Since \(\rho_{\pm}\) behave as projectors,
one halves of \(\theta^{\alpha}\) and \(\widehat{\theta}^{\hat{\alpha}}\)
are absent from the equations of motion.

\paragraph{Pure spinor}

Equations of motion for the fermionic fluctuations \(X^{1}\) and \(X^{3}\) are
\begin{align}
[D_{+},[D_{-},X^{1}]] + [J_{-}^{2},[D_{+},X^{3}]]  &= 0,\quad
[D_{-},[D_{+},X^{3}]] + [J_{+}^{2},[D_{-},X^{1}]]  = 0
\end{align}
or in the component notation
\begin{align}
\label{eqn:PSflucfermi}
D_{+}(D_{-}\theta)^{\alpha} - {1\over2}(\eta\rho_{-})^{\alpha}{}_{\hat{\beta}}(D_{+}\widehat{\theta})^{\hat{\beta}} & = 0,\quad
D_{-}(D_{+}\widehat{\theta})^{\hat{\alpha}} + {1\over2}(\eta\rho_{+})^{\hat{\alpha}}{}_{\beta}(D_{-}\theta)^{\beta}  = 0
\mathperiod
\end{align}
Note well the difference and resemblance of these to the
corresponding equations in the Green-Schwarz formalism~(\ref{eqn:GSflucfermi}).
Unlike in the Green-Schwarz formalism,
equations of motion~(\ref{eqn:PSflucfermi}) for fermionic fluctuations
here are of second order and non-degenerate.
On the other hand, if one defines
\(S=(\eta\rho_{-}\theta)\) and \(\widehat{S}=(\eta\rho_{+}\widehat{\theta})\) in Green-Schwarz formalism
and \(\Theta=(D_{-}\theta)\) and \(\widehat{\Theta}=(D_{+}\widehat{\theta})\) in pure spinor formalism,
the equations here can be obtained by formally replacing
\((S,\widehat{S})\) in~(\ref{eqn:GSflucfermi}) by \((\Theta,\widehat{\Theta})\).
Since \((\Theta,\widehat{\Theta})\) do not contain the projectors \(\rho_{\pm}\) as \((S,\widehat{S})\) do,
one cannot immediately identify them with \((S,\widehat{S})\),
but we shall show in the subsection~\ref{subsec:fermifluc}
that one can further split \((\Theta,\widehat{\Theta})\)
to the Green-Schwarz like degrees of freedom \((S,\widehat{S})\) and the rest,
at least around the classical solutions contained in an \(\mathbb{R}_{t}\times S^{2}\subset AdS_{5}\times S^{5}\).

Equations of motion for the pure spinor ghosts are simply
\begin{align}
[D_{-},\lambda^{1}] &= [D_{-},w^{3}] = 0,\quad
[D_{+},\widehat{\lambda}^{3}] = [D_{+},\widehat{w}^{1}] = 0
\end{align}
or
\begin{align}
D_{-}\lambda^{\alpha} &= D_{-}w_{\alpha} = 0,\quad
D_{+}\widehat{\lambda}^{\hat{\alpha}} = D_{+}\widehat{w}_{\hat{\alpha}} = 0
\mathperiod
\end{align}
Note that \((D_{-}\theta)=(D_{+}\widehat{\theta})=0\) is a solution
to the equations of motion~(\ref{eqn:PSflucfermi}).
So there are \(22\times2\) bosonic modes and \(16\times2\) fermionic modes
satisfying the same equations of motion,
and one already expects a huge cancellation of zero-point energies.

\subsubsection{BRST transformations of fluctuations}

Although we will not need it in this article,
the action of the BRST symmetry on fluctuations \(X=X^{1}+X^{2}+X^{3}\) can be computed from
the ``finite'' BRST transformation
\begin{align}
g = \underline{g}\mathe^{X} \to \underline{g}\mathe^{X}\mathe^{\lambda^{1} + \widehat{\lambda}^{3}}
\end{align}
by using the Baker-Campbell-Hausdorff formula.
To the second order in fluctuations, they are given by
\begin{align}
\begin{split}
\delta_{\text{B}} X^{2} &= 0 + {1\over2}([X^{1},\lambda^{1}] + [X^{3},\widehat{\lambda}^{3}]) + \cdots \mathcomma\\
\delta_{\text{B}} X^{1} &= \lambda^{1} + {1\over2}[X^{2},\lambda^{3}] + \cdots,\quad
\delta_{\text{B}} X^{3} = \lambda^{1} + {1\over2}[X^{2},\lambda^{3}] + \cdots
\mathperiod
\end{split}
\end{align}
Note that, because of pure spinor constraints \(\{\lambda^{1},\lambda^{1}\}=\{\widehat{\lambda}^{3},\widehat{\lambda}^{3}\}=0\),
the right hand sides of these equations are linear in \((\lambda^{1},\widehat{\lambda}^{3})\).
Pure spinors \(\lambda^{1}\) and \(\widehat{\lambda}^{3}\)
are BRST invariant and the conjugates \(w^{3}\) and \(\widehat{w}^{1}\) transform as
\begin{align}
\delta_{\text{B}} w^{3} &= -[D_{+},X^{3}] - [\underline{J}^{2}_{+},X^{1}] + \cdots,\quad
\delta_{\text{B}} \widehat{w}^{1} = -[D_{-},X^{1}] - [\underline{J}^{2}_{+},X^{3}] + \cdots
\mathperiod
\end{align}

\subsection{A family of classical solutions in \texorpdfstring{$AdS_{5}\times S^{5}$}{AdS5 x S5}}
\label{subsec:family}

For simplicity, we from now on restrict ourselves to a rather simple family of classical solutions
in which the string sits at the center of \(AdS_{5}\) and (possibly) extended
in an \(S^{2}\subset S^{5}\). Moreover, we assume that the string is rigid,
meaning that the coefficients of fluctuation action is \(\tau\)-independent.\footnote{%
The rigidity assumption is for facilitating the proof of a relation between
the one-loop correction to spacetime energy and the expectation value of worldsheet Hamiltonian
(see next subsection);
it is unnecessary for the comparison of semiclassical partition functions
of the Green-Schwarz and pure spnior formalisms.}
More concretely,
if one denotes \(AdS\) time by \(t\) and
azimuthal and polar angles of \(S^{2}\) by \((\psi,\phi)\)
with \(\psi=0\sim \pi\) and \(\phi=0\sim2\pi\),
a solution in the family can be written as
\begin{align}
\label{eqn:class}
t &= \kappa \tau,\quad
\psi = \psi(\sigma),\quad
\phi = \nu \tau + \phi_{0}(\sigma)
\end{align}
for some constants \(\kappa\) and \(\nu\),
and \(\tau\)-independent functions \(\psi(\sigma)\) and \(\phi_{0}(\sigma)\).
Solutions in this class include
the point-like rotating BMN string~\cite{Berenstein:2002jq},
the folded spinning string~\cite{Gubser:2002tv},
and if the periodicity in \(\sigma\) direction is relaxed,
the giant magnon~\cite{Hofman:2006xt}.

We shall identify \((t,\psi,\phi)\) directions to the directions generated by \((T_{0},T_{8},T_{9})\in\mathfrak{g}^{2}\).
The parameterization of the coset representative \(g(\tau,\sigma)\)
in terms of \((t,\psi,\phi)\) is then
\begin{align}
g &= \mathe^{t T_{0}}\mathe^{\phi T_{9}}\mathe^{(\psi-\pi/2)T_{8}}
\mathperiod
\end{align}
The non-vanishing components of the Metsaev-Tseytlin current are
\begin{align}
J_{\pm}
&\equiv g^{-1}\partial_{\pm}g
{} = \partial_{\pm}t T_{0}  + \partial_{\pm}\psi T_{8} + \partial_{\pm} \phi \sin\psi T_{9} - \partial_{\pm} \phi\cos \psi T_{89}
\mathperiod
\end{align}
Components of the current \(J_{\pm}\) are just the pullbacks of
vielbein and spin connection on \(S^{2}\)
\begin{align}
e^{0}_{t} &=1,\quad
e^{8}_{\psi} =1,\quad
e^{9}_{\phi} =\sin \psi
\mathcomma\\
\omega_{\phi}{}^{89} &= \cos\psi
\mathperiod
\end{align}

\subsection{Relation between \texorpdfstring{$\Delta E$}{dE} and worldsheet Hamiltonian \texorpdfstring{$H_{2}$}{H2}}
\label{subsec:relEH}

For the class of solutions described in the previous subsection,
the one-loop correction to the spacetime energy
\(\langle \Omega| (E-\underline{E}) |\Omega\rangle\)
has a rather simple relation to a properly defined worldsheet Hamiltonian \(H_{2}\) for fluctuations.
This is well-known in the Green-Schwarz formalism (both in conformal and static gauges)
and it will be shown here that the same is true for the pure spinor formalism as well.
To be more specific, it will now be shown that the relation\footnote{%
Here, \((J,\underline{J})\) are an angular momentum in \(S^{5}\)
and its classical value, and have nothing to do with the Metsaev-Tseytlin current \(J\).}
\begin{align}
\label{eqn:relation}
\langle \Psi| \bigl( \kappa(E-\underline{E}) - \nu(J-\underline{J}) \bigr) |\Psi\rangle
&= \langle \Psi| H_{2} |\Psi\rangle
\end{align}
holds for any quantum state \(|\Psi\rangle\) in the BRST cohomology built on the ground state \(|\Omega\rangle\).
Moreover, since \(J\) is a compact generator with discrete eigenvalues,
the ground state \(|\Omega\rangle\) is supposed to have the same eigenvalue \(\underline{J}\)
as the classical solution.
Exploiting the relation~(\ref{eqn:relation}) is useful
because the expectation value of \(H_{2}\) (zero-point energy)
for the ground state
\(|\Omega\rangle\) can be computed from the one-loop partition function of fluctuations.

A proof of a relation of the type~(\ref{eqn:relation})
in the Green-Schwarz formalism in a conformal gauge is given~\cite{Frolov:2002av} by noting
\begin{align}
\label{eqn:GSEvsH}
\kappa(E-\underline{E}) - \nu(J-\underline{J}) + (L_{0} +\widebar{L}_{0}) \approx H_{2}
\end{align}
where \(L_{0}+\widebar{L}_{0}\) is the zero-mode of the Green-Schwarz Virasoro operator
(including contributions from reparameterization ghosts)
expanded to quadratic order in fluctuations
and the equality holds up to fermionic constraints of the Green-Schwarz formalism.
In~(\ref{eqn:GSEvsH})
both \(\kappa (E-\underline{E}) -  \nu (J-\underline{J})\) and \(L_{0}+\widebar{L}_{0}\) contain
terms linear in fluctuations along a lightcone direction,
but the linear terms cancel in the sum
and the remaining expression quadratic in fluctuations coincides with \(H_{2}\).
In simple situations where one can take a lightcone gauge,
the Hamiltonian \(H_{2}\) can be decomposed into three pieces
\(H_{\text{phys}} + H_{\text{lc}} + H_{bc}\)
each representing the Hamiltonian for physical transverse directions,
lightcone directions (\(x^{\pm} = t\pm \phi\)),
and reparameterization ghosts.
Contributions from \(H_{\text{lc}} + H_{bc}\) cancel out
from the expectation value \(\langle \Psi | H_{2} | \Psi \rangle\)
in the right hand side of~(\ref{eqn:relation})
and leaves a result identical to the one in a lightcone gauge.

In the pure spinor formalism, even though the Virasoro operator is not a constraint,
a cohomology of the BRST operator
has to have a vanishing eigenvalue of \(L_{0}+\widebar{L}_{0}\)
since there is a composite \(b\)-ghost that makes the Virasoro operator trivial.
So one hopes that the expression of the form~(\ref{eqn:relation})
with \(L_{0}+\widebar{L}_{0} = \{\QB,b_{0}+\widebar{b}_{0}\}\)
is also true in the pure spinor formalism.
Although the appearance of Virasoro operators
as well as the charges \((E,J)\) and Hamiltonians
in Green-Schwarz and pure spinor formalisms
are quite different, this hope turns out to be true.

The rest of this subsection is devoted to some details of the proof of~(\ref{eqn:relation}).
First we note that the proper definition of the quadratic Hamiltonian
(written in terms of ``velocity variables'') should be
\begin{align}
\label{eqn:H2}
H_{2} &= H_{2B} + H_{2F} + H_{2G} = {1\over2\pi\alpha'}\int\mathd\sigma (\calH_{2B} + \calH_{2F} + \calH_{2G})
\end{align}
where
\begin{align}
\begin{split}
\calH_{2B}
&= {1\over4}\str\Bigl(
([\partial_{\tau},X^{2}])^{2}
{} - ([\underline{J}^{0}_{\tau},X^{2}])^{2}
{} + ([\underline{J}^{2}_{\tau},X^{2}])^{2}
{} + ([D_{\sigma},X^{2}])^{2}
{} - ([\underline{J}^{2}_{\sigma},X^{2}])^{2}
\Bigr)
\mathcomma
\\
\calH_{2F}
&= \str\Bigl( [D_{+} - \underline{J}^{0}_{\tau},X^{1}][D_{+},X^{3}]
{} - [\underline{J}^{0}_{-},X^{1}][\underline{J}^{2}_{+},X^{1}]
{} + {1\over2}[\partial_{\sigma},X^{1}][\underline{J}^{2}_{\tau},X^{1}] \nonumber\\
&{}\qquad + [D_{-} - \underline{J}^{0}_{\tau},X^{3}][D_{-},X^{1}]
{} - [\underline{J}^{0}_{+},X^{3}][\underline{J}^{2}_{-},X^{3}]
{} - {1\over2}[\partial_{\sigma},X^{3}][\underline{J}^{2}_{\tau},X^{3}]\Bigr)
\mathcomma
\\
\calH_{2G} &= \str\Bigl(w^{3}[D_{+},\lambda^{1}] + \widehat{w}^{1}[D_{-},\widehat{\lambda}^{3}]
{} - N\underline{J}^{0}_{\tau} - \widehat{N}\underline{J}^{0}_{\tau}\Bigr)
\mathperiod
\end{split}
\end{align}
The bosonic Hamiltonian \(H_{2B}\) is nothing but the canonical Hamiltonian
computed from the quadratic Lagrangian \(L_{2B}\) of~(\ref{eqn:GSS2B}),
\begin{align}
H_{2B} &= P_{2}\partial_{\tau}X^{2} - L_{2B}, \quad
P_{2} \equiv {\partial L_{2B} \over \partial (\partial_{\tau}X^{2})} = {1\over4\pi\alpha'}[D_{\tau},X^{2}]
\mathperiod
\end{align}
The Hamiltonians for fermions \(H_{2F}\) and ghosts \(H_{2G}\) are
not in a naive canonical form,
but they reduce to the standard Hamiltonians for the second order fermions
and the left and right moving \(\beta\gamma\) systems of weight \((1,0)\)
when the coupling to the background currents \(\underline{J}^{0}\) and \(\underline{J}^{2}\)
is dropped.
The coupling to the background currents is fixed by the BRST symmetry
up to an addition of BRST trivial terms
so we claim that~(\ref{eqn:H2}) is the correct Hamiltonian for the quadratic fluctuations.

As mentioned above,
in order to relate the one-loop correction to the spacetime energy
to the expectation value of \(H_{2}\),
it is convenient to look at the quantity
\begin{align}
\begin{split}
\kappa E - \nu J
&= -{1\over4\pi\alpha'}\int\mathd\sigma\str\Bigl(j_{\tau}\bigl( (\partial_{\tau}\underline{t})T_{0} + (\partial_{\tau}\underline{\phi})T_{9} \bigr)\Bigr)
\end{split}
\end{align}
where
\begin{align}
\label{eqn:jtau}
j_{\tau}=j_{+}+j_{-}=g(J^{2}_{\tau}+J^{1}_{\tau}+J^{3}_{\tau}-{1\over2}(J^{1}_{\sigma}-J^{3}_{\sigma}) + 2N + 2\widehat{N})g^{-1}
\end{align}
is the \(\tau\)-component of
the \(PSU(2,2|4)\) Noether current defined in~(\ref{eqn:NoetherCurrent}).
Classical values \((\underline{E},\underline{J})\)
of \((E,J)\) are given by
\begin{align}
\underline{E} &= {1\over4\pi\alpha'}\int\mathd \sigma \partial_{\tau}\underline{t} = {\kappa\over2\alpha'},\quad
\underline{J} = {1\over4\pi\alpha'}\int\mathd \sigma \sin^{2}\underline{\psi} \partial_{\tau}\underline{\phi}
{} = {\nu\over4\pi\alpha'}\int\mathd \sigma \sin^{2}\underline{\psi}
\end{align}
and semiclassical expressions for \((E,J)\) can be computed
by separating \(g=g(\tau,\sigma)\) and the currents \((J_{\tau},J_{\sigma},N,\widehat{N})\)
in~(\ref{eqn:jtau}) to their background values and fluctuations.
(Recall that we are expanding around a trivial ghost profile
so \(N\) and \(\widehat{N}\) are understood to be quadratic in fluctuations.)
It is useful to note that the rigidity assumption \(\partial_{\tau}\underline{\psi}=0\) implies
\begin{align}
\underline{g}^{-1}((\partial_{\tau}\underline{t})T_{0} + (\partial_{\tau}\underline{\phi})T_{9})\underline{g}
&= (\partial_{\tau}\underline{t}T_{0} + \partial_{\tau}\underline{\phi}\sin\underline{\psi}T_{9}) - \partial_{\tau}\underline{\phi}\cos\underline{\psi}T_{89}
{} = \underline{J}^{2}_{\tau} + \underline{J}^{0}_{\tau}
\mathperiod
\end{align}
Computation of \(\kappa E - \nu J\) is then straightforward
and to the quadratic order in fluctuations it is given by
\begin{align}
\kappa E - \nu J
&= \kappa\underline{E} - \nu\underline{J} - {1\over2\pi\alpha'}\int\mathd\sigma(\calC_{1} + \calC_{2B} + \calC_{2F}+ \calC_{2G})
\end{align}
where
\begin{align}
\begin{split}
\calC_{1} &= {1\over2}\str\Bigl(([D_{\tau},X^{2}] + [\underline{J}^{0}_{\tau},X^{2}])\underline{J}^{2}_{\tau}\Bigr)
\mathcomma\\
\calC_{2B} &= {1\over2}\str\Bigl([D_{\tau},X^{2}][\underline{J}^{0}_{\tau}, X^{2}] - ([\underline{J}^{2}_{\tau},X^{2}])^{2}\Bigr)
\mathcomma\\
\calC_{2F} &= {1\over4}\str\Bigl(
{} - [D_{\sigma},X^{1}][\underline{J}^{2}_{\tau},X^{1}]
{} + 2[\underline{J}^{0}_{\tau},X^{1}][\underline{J}^{2}_{\tau},X^{1}]
{} + [\underline{J}^{0}_{\tau},X^{1}][\underline{J}^{2}_{\sigma},X^{1}] \nonumber\\
{}&\quad
{} + [D_{\sigma},X^{3}][\underline{J}^{2}_{\tau},X^{3}]
{} + 2[\underline{J}^{0}_{\tau},X^{3}][\underline{J}^{2}_{\tau},X^{3}]
{} - [\underline{J}^{0}_{\tau},X^{3}][\underline{J}^{2}_{\sigma},X^{3}]\nonumber\\
{}&\quad - [\underline{J}^{2}_{\tau},X^{1}][\underline{J}^{2}_{\sigma},X^{3}]
{} + [\underline{J}^{2}_{\sigma},X^{1}][\underline{J}^{2}_{\tau},X^{3}]
{} + 2[D_{\tau},X^{1}][\underline{J}^{0}_{\tau}, X^{3}]
{} + 2[\underline{J}^{0}_{\tau}, X^{1}][D_{\tau},X^{3}] \nonumber\\
{}&\quad - [D_{\sigma},X^{1}][\underline{J}^{0}_{\tau},X^{3}]
{} + [\underline{J}^{0}_{\tau},X^{1}][D_{\sigma},X^{3}]
\Bigr)
\mathcomma\\
\calC_{2G} &= \str\Bigl( (N + \widehat{N})\underline{J}^{0}_{\tau} \Bigr)
\mathperiod
\end{split}
\end{align}

The semiclassical expression for the worldsheet energy \(L_{0}+\widebar{L}_{0}\)
(which is BRST trivial) can be computed in a similar manner.
To quadratic order in fluctuations it is given by
\begin{align}
L_{0}+\widebar{L}_{0}
&= {1\over2\pi\alpha'}\int\mathd\sigma (\calL_{1} + \calL_{2B} + \calL_{2F} + \calL_{2G})
\end{align}
where
\begin{align}
\begin{split}
\calL_{1}
&= \str(\underline{J}^{2}_{+}[D_{+},X^{2}] + \underline{J}^{2}_{-}[D_{-},X^{2}])
\mathcomma\\
\calL_{2B}
&= {1\over2}\str\Bigl( ([D_{+},X^{2}])^{2} + ([D_{-},X^{2}])^{2} - ([\underline{J}^{2}_{+},X^{2}])^{2} - ([\underline{J}^{2}_{-},X^{2}])^{2} \Bigr)
\mathcomma\\
\calL_{2F}
&= \str\Bigl( [D_{+},X^{1}][D_{+},X^{3}] + [D_{-},X^{1}][D_{-},X^{3}] \nonumber\\
&\qquad + {1\over2}\sum_{i=1,3}([\underline{J}^{0}_{+},X^{i}][\underline{J}^{2}_{+},X^{i}]
{} + [\underline{J}^{0}_{-},X^{i}][\underline{J}^{2}_{-},X^{i}])
\Bigr)
\mathcomma
\\
\calL_{2G} &= \str\Bigl( w[D_{+},\lambda] + \widehat{w}[D_{-},\widehat{\lambda}]\Bigr)
\mathperiod
\end{split}
\end{align}

Upon integrating a \(\sigma\)-derivative by parts
and using the Maurer-Cartan equation
as well as \(\partial_{\tau}\underline{J}^{2}_{\mu}=0\)
(the rigidity assumption on the classical solution),
\(\calL_{1}\) is found to be equal to \(\calC_{1}\).
Then, one finds that the sum of 
\(\kappa(E-\underline{E})-\nu(J-\underline{J})\) and \((L_{0}+\widebar{L}_{0})\)
only contains terms quadratic in fluctuations
and is nothing but the worldsheet Hamiltonian \(H_{2}\):
\begin{align}
\kappa(E-\underline{E})-\nu(J-\underline{J}) + (L_{0}+\widebar{L}_{0}) &= H_{2}
\mathperiod
\end{align}
This is the analogue of~(\ref{eqn:GSEvsH}) for the pure spinor formalism
that we wanted to show.
Note that this incidentally shows that \(H_{2}\) is BRST invariant,
since both \(PSU(2,2|4)\) and Virasoro charges are BRST invariant.

\subsection{Disentangling fermionic fluctuations}
\label{subsec:fermifluc}

Here, we study in detail the fermionic fluctuations around the family of
classical solutions~(\ref{eqn:class}) but with the rigidity assumption relaxed:
\begin{align*}
t &= \kappa \tau,\quad
\psi = \psi(\tau,\sigma),\quad
\phi = \phi(\tau,\sigma),\quad (\psi,\phi)\in S^{2}\subset S^{5}
\mathperiod
\end{align*}
For notational simplicity we set \(\kappa = 2\alpha'\underline{E} = 1\) by adjusting \(\alpha'\).

\paragraph{Green-Schwarz}

We first study the fermionic fluctuations \((\theta^{\alpha},\widehat{\theta}^{\hat{\alpha}})\)
in the Green-Schwarz formalism whose equations of motion are~(\ref{eqn:GSflucfermi})
\begin{align*}
D_{+}(\eta \rho_{-}\theta)^{\alpha} - {1\over2}(\eta\rho_{-})^{\alpha}{}_{\hat{\beta}}(\eta\rho_{+}\widehat{\theta})^{\hat{\beta}} &= 0\,,\quad
D_{-}(\eta \rho_{+}\widehat{\theta})^{\hat{\alpha}} + {1\over2}(\eta\rho_{+})^{\hat{\alpha}}{}_{\beta}(\eta\rho_{-}\theta)^{\beta} = 0
\mathperiod
\end{align*}
For the class of solutions at hand, matrices \(\rho_{\pm}\)
and covariant derivatives \(D_{\pm}\) can be diagonalized neatly.

It will be convenient to take our basis of \(16\times 16\) \(\gamma\)-matrices to have
\begin{align}
\gamma_{8} = (-\sigma_{2}\otimes 1_{8}),\quad
\gamma_{9} = (\sigma_{1}\otimes 1_{8})
\end{align}
so that the spin connection becomes diagonal:
\begin{align}
\omega_{\pm} &= \partial_{\pm}\phi \, \omega_{\phi}^{89}\gamma_{89},\quad
\gamma_{89} \equiv {1\over2}(\gamma_{8}\gamma_{9}-\gamma_{9}\gamma_{8}) = {\mathi\over2}(\sigma_{3}\otimes 1_{8})
\mathperiod
\end{align}
Below, we shall often display \(\gamma\)-matrices in a \(2\times 2\) format
and leave the trivial factor of \(1_{8}\) implicit.
In this basis, \(\rho_{\pm}=\partial_{\pm}x^{m}e_{m}^{a}\gamma_{a}\) takes the form
\begin{align}
(\rho_{\pm})_{\alpha\beta}
{} = \gamma_{0} + a_{\pm}(-\sigma_{2}\otimes 1_{8}) + b_{\pm}(\sigma_{1}\otimes 1_{8})
{} = \begin{pmatrix}1 & c_{\pm} \\ c_{\pm}^{\ast} & 1\end{pmatrix}
\end{align}
where
\begin{align}
a_{\pm} &= \partial_{\pm}\underline{\psi},\quad b_{\pm} = \partial_{\pm}\underline{\phi}\sin\underline{\psi},\quad
c_{\pm} = \mathi a_{\pm} + b_{\pm}
\mathperiod
\end{align}
Note that the Virasoro condition implies
\begin{align}
c_{\pm}^{\ast}c_{\pm} = a_{\pm}^{2} + b_{\pm}^{2} = 1
\end{align}
so \(c_{\pm}\) are complex numbers of modulus \(1\).
We denote by \(\alpha_{\pm}\) the phase of \(c_{\pm}\):
\begin{align}
c_{\pm} &= \mathe^{\mathi \alpha_{\pm}}
\mathperiod
\end{align}
Classical equations of motion for the background field implies
\begin{align}
(\partial_{\pm}-\mathi \omega_{\pm})c_{\mp} = 0,\quad
\partial_{\pm} \alpha_{\mp} = \omega_{\pm}
\mathperiod
\end{align}

With these notational preparation,
it is straightforward to find a basis in which
\(\rho_{\pm}\) and \(D_{\pm}\) simplify simultaneously.
Namely, for
\begin{align}
\label{eqn:matUV}
U &= \begin{pmatrix}
{}    \mathe^{-{\mathi\over2} \alpha_{+}} & \mathe^{+{\mathi\over2} \alpha_{+}} \\
{}    -\mathe^{-{\mathi\over2} \alpha_{+}} &\mathe^{+{\mathi\over2}\alpha_{+}}
{} \end{pmatrix},\quad
V = \begin{pmatrix}
{}    \mathe^{-{\mathi\over2} \alpha_{-}} & \mathe^{+{\mathi\over2} \alpha_{-}} \\
{}    -\mathe^{-{\mathi\over2} \alpha_{-}} &\mathe^{+{\mathi\over2}\alpha_{-}}
{} \end{pmatrix}
\end{align}
one finds that
\begin{align}
\rho_{+} &= U^{-1}\begin{pmatrix}2&0\\0&0\end{pmatrix}U,\quad
\rho_{-} = V^{-1}\begin{pmatrix}2&0\\0&0\end{pmatrix}V
\mathcomma\\
D_{+} &= V^{-1}\partial_{+}V \equiv \partial_{+} +  V^{-1}(\partial_{+}V),\quad
D_{-} = U^{-1}\partial_{-}U \equiv \partial_{-} +  U^{-1}(\partial_{+}U)
\mathperiod
\end{align}
Substituting these into the equations of motion, one finds
\begin{align}
\label{eqn:GSmatrixeq}
\begin{split}
\partial_{+}\begin{pmatrix}1 & 0 \\ 0 & 0\end{pmatrix} V\theta
{} - \eta\begin{pmatrix}1&0 \\ 0 & 0\end{pmatrix}VU^{-1}
{} \begin{pmatrix}1 & 0 \\ 0 & 0\end{pmatrix} U\widehat{\theta} &= 0
\mathcomma\\
\partial_{-}\begin{pmatrix}1 & 0 \\ 0 & 0\end{pmatrix} U\widehat{\theta}
{} + \eta\begin{pmatrix}1&0 \\ 0 & 0\end{pmatrix}UV^{-1}
{} \begin{pmatrix}1 & 0 \\ 0 & 0\end{pmatrix} V\theta &= 0
\mathperiod
\end{split}
\end{align}
This clearly shows that one-halves of \(V\theta\) and \(U\widehat{\theta}\) do not propagate.

To be more concrete, introduce variables \((S,\widehat{S},T,\widehat{T})\) and \(\beta\) via
\begin{align}
V\theta &= \begin{pmatrix}S \\ T\end{pmatrix},\quad
U\widehat{\theta} = \begin{pmatrix}\widehat{S} \\ \widehat{T}\end{pmatrix}
\mathcomma\\
\beta &= {1\over2}(\alpha_{+} - \alpha_{-})\;\to\;
UV^{-1} = \begin{pmatrix}\cos \beta & \mathi \sin \beta \\ \mathi \sin \beta & \cos \beta\end{pmatrix}
\mathperiod
\end{align}
Then, \(T\) and \(\widehat{T}\) decouple from the equations of motion and
\(S\) and \(\widehat{S}\) obey
\begin{align}
\label{eqn:GSeqS}
\nabla_{\text{GS}}
\begin{pmatrix}
S \\ \widehat{S}
\end{pmatrix} &= 0,\quad
\nabla_{\text{GS}} \equiv
\begin{pmatrix}
\partial_{+} & -\eta\cos \beta \\
\eta\cos \beta & \partial_{-}
\end{pmatrix}
\mathperiod
\end{align}

It is amusing to note that the combination
\(\phi_{s} = 2\beta =(\alpha_{+}-\alpha_{-})\) is the solution to the sine-Gordon equation
\(4\partial_{+}\partial_{-}(\phi_{s})=\sin(\phi_{s})\)
which determines our solution \((t,\psi,\phi)\) completely.
For example, \(\beta=0\) corresponds to the rotating point-like string
\(t = \phi = \kappa \tau,\; \psi=\pi/2\) of~\cite{Berenstein:2002jq}
and~(\ref{eqn:GSeqS}) reduces to the well-known equations of motion
for the lightcone fermions in a Ramond-Ramond plane-wave background~\cite{Metsaev:2002re}.

\paragraph{Pure spinor}

Recall that the coupled equations of motion for fluctuations are
\begin{align}
\label{eqn:coupledPSfermi2}
D_{+}(D_{-}\theta)^{\alpha} - {1\over2}(\eta \rho_{-})^{\alpha}{}_{\hat{\beta}}(D_{+}\widehat{\theta})^{\hat{\beta}} & = 0,\quad
D_{-}(D_{+}\widehat{\theta})^{\hat{\alpha}} + {1\over2}(\eta \rho_{+})^{\hat{\alpha}}{}_{\beta}(D_{-}\theta)^{\beta} = 0
\mathperiod
\end{align}
These have two ``branches'' of solutions.
First branch is given by
\begin{align}
D_{-}\theta = D_{+}\widehat{\theta} = 0
\end{align}
where \(D_{-}\theta=0\) implies \(D_{+}\widehat{\theta}=0\) and vice versa.
To show that \(D_{-}\theta=0\) implies \(D_{+}\widehat{\theta}=0\),
denote for convenience
\begin{align}
\Psi &= U\theta,\quad \widehat{\Psi} = V\widehat{\theta}
\mathperiod
\end{align}
Note that \(D_{-}\theta = 0\) is equivalent to \(\partial_{-}\Psi = 0\)
and that \(D_{+}\widehat{\theta} = 0\) is equivalent to \(\partial_{+}\widehat{\Psi} = 0\).
Now, assuming \(\partial_{-}\Psi = 0\), the equations of~(\ref{eqn:coupledPSfermi2})
imply that \(\widehat{\Psi}\) satisfies
\begin{align}
{}\begin{pmatrix}1&0\\0&0\end{pmatrix}\partial_{+}\widehat{\Psi}
{} = 0,\quad
{} \partial_{-}\begin{pmatrix}\cos \beta & \mathi\sin \beta \\ \mathi\sin \beta & \cos \beta\end{pmatrix}\partial_{+}\widehat{\Psi}
{} = 0
\mathperiod
\end{align}
In terms of the \(8+8\) splitting \(\widehat{\Psi} = \begin{pmatrix}\widehat{\Psi}_{1} \\ \widehat{\Psi}_{2}\end{pmatrix}\)
these are equivalent to
\begin{align}
\partial_{+}\widehat{\Psi}_{1} &= 0,\quad
\partial_{+}\partial_{-}\widehat{\Psi}_{2} = -(\partial_{-}\beta)\cot\beta\partial_{+}\widehat{\Psi}_{2},\quad
\partial_{+}\partial_{-}\widehat{\Psi}_{2} = (\partial_{-}\beta)\tan \beta\partial_{+}\widehat{\Psi}_{2}
\mathperiod
\end{align}
Thus for non-constant \(\beta\)
one finds \(\partial_{+}\widehat{\Psi}_{2}=0\) as well.
When \(\beta\) is a constant its only possible values are \(0\bmod \pi/2\)
since \(2\beta\) is a solution to the sine-Gordon equation.
Then equations of motion for \(\widehat{\Psi}_{2}\) is just
\(\partial_{+}\partial_{-}\widehat{\Psi}_{2}=0\) and one can include a half of the solutions
\(\partial_{+}\widehat{\Psi}_{2}=0\) in the present branch,
and the other half \(\partial_{-}\widehat{\Psi}_{2}=0\) in the other branch described shortly.
This completes the proof that \(D_{-}\theta=0\) implies \(D_{+}\widehat{\theta}=0\),
and we have learnt that this solution branch consists of
\(16\) left-moving fields \(\Psi^{\alpha}(\sigma^{-})\)
and \(16\) right-moving fields \(\widehat{\Psi}^{\hat{\alpha}}(\sigma^{+})\).

To describe the other branch,
it is useful to introduce the variables \((S,\widehat{S},T,\widehat{T})\) via
\begin{align}
V(D_{-}\theta) &= \begin{pmatrix}{S} \\ {T}\end{pmatrix},\quad
U(D_{+}\widehat{\theta}) = \begin{pmatrix}\widehat{S} \\ \widehat{T}\end{pmatrix}
\mathperiod
\end{align}
Since we have already taken care of the branch \(D_{-}\theta=D_{+}\widehat{\theta}=0\),
one may assume that neither \((S,T)\) nor \((\widehat{S},\widehat{T})\) is identically zero.
Equations of motion for \((S,\widehat{S},T,\widehat{T})\) are found to be
\begin{align}
\begin{split}
\partial_{+}\begin{pmatrix}S \\ T \end{pmatrix}
{} - \eta\begin{pmatrix}1&0\\0&0\end{pmatrix}\begin{pmatrix}\cos \beta& -\mathi \sin\beta \\ -\mathi \sin \beta& \cos \beta\end{pmatrix}
{} \begin{pmatrix}\widehat{S} \\ \widehat{T} \end{pmatrix} &= 0
\mathcomma\\
\partial_{-}\begin{pmatrix}\widehat{S} \\ \widehat{T} \end{pmatrix}
{} + \eta\begin{pmatrix}1&0\\0&0\end{pmatrix}\begin{pmatrix}\cos \beta& \mathi \sin\beta \\\mathi \sin \beta& \cos \beta\end{pmatrix}
{} \begin{pmatrix}S \\ T \end{pmatrix} &= 0
\mathperiod
\end{split}
\end{align}
Compared to the Green-Schwarz equations in the same basis~(\ref{eqn:GSmatrixeq}),
one here does not have projections to \((S,\widehat{S})\)
so there remains a mixing between \((S,\widehat{S})\)
and \((T,\widehat{T})\):
\begin{align}
\label{eqn:PSmatrixEQ}
\nabla_{F}\begin{pmatrix}
S \\ \widehat{S} \\ T \\ \widehat{T}
\end{pmatrix} &= 0,\quad
\nabla_{F} \equiv
\begin{pmatrix}
\partial_{+} & -\eta\cos \beta & 0 & \mathi \eta\sin\beta \\
\eta\cos \beta & \partial_{-} & \mathi \eta\sin\beta & 0 \\
0 & 0 & \partial_{+} & 0 \\
0 & 0& 0& \partial_{-}
\end{pmatrix}
\mathperiod
\end{align}
However, the mixing is minor as can be seen from the
block triangular structure of the matrix differential operator \(\nabla_{F}\) in~(\ref{eqn:PSmatrixEQ}).
In particular, equations of motion for \(T\) and \(\widehat{T}\) are simply
\(\partial_{+}T=\partial_{-}\widehat{T}=0\) and the functional determinant
of \(\nabla_{F}\) factorize as
\begin{align}
\label{eqn:factorization}
\det\nabla_{F}  = (\det\nabla_{\text{GS}})(\det\partial_{+})(\det\partial_{-})
\end{align}
where \(\det\nabla_{\text{GS}}\) is the functional determinant of
the matrix differential operator appeared in the
equations of motion~(\ref{eqn:GSeqS}) for the Green-Schwarz fermions.
Although we do not quite pretend to have shown the factorization~(\ref{eqn:factorization}) rigorously,
we believe that it is possible to do so
for example by employing the technique of~\cite{Beccaria:2010ry}.

\subsection{Comparison of 1-loop corrections}
\label{subsec:cmp1loop}

\paragraph{Partition function}

Based on the analyses made thus far,
it will now be shown that one-loop partition functions of fluctuations
in Green-Schwarz and pure spinor formalisms agree
for any classical solution in the family of subsection~\ref{subsec:family}.
The following table summarizes the contributions of various fluctuations
to the partition functions
(the partition function for the Green-Schwarz formalism
is for a conformal semilightcone gauge in which non-propagating
fermionic fluctuations are dropped):
\begin{small}
\begin{center}
\begin{tabular}{l|ccccc}
{} &  Bosons & Fermions && Ghosts &\\ \hline
Green-Schwarz & \(t,\psi,\phi,\;\; x^{i}\) & \(S^{A},\widehat{S}^{A}\) & -- & \(b,c,\widebar{b},\widebar{c}\) & -- \\[1ex]
{(conf.\ gauge)} & \((\det \Delta_{3})^{-1}(\det\Delta_{7})^{-1}\) & \(\det\nabla_{\text{GS}}\) & {--} & {\((\det\Box)^{2}\)} & {--}
\\[1.5ex] \hline
Pure spinor & \(t,\psi,\phi,\;\; x^{i}\) & \(S^{A},\widehat{S}^{A}\) & \(T^{\dot{A}},\widehat{T}^{\dot{A}},\Psi^{\alpha},\widehat{\Psi}^{\hat{\alpha}}\)  & -- & \(w_{\alpha},\lambda^{\alpha},\widehat{w}_{\dot{\alpha}},\widehat{\lambda}^{\alpha}\)\\[1ex]
{} & \((\det \Delta_{3})^{-1}(\det\Delta_{7})^{-1}\) & {\(\det\nabla_{\text{GS}}\)} &{\((\det\Box)^{8+16}\)} & {--} & {\((\det\Box)^{-22}\)}
\end{tabular}

\medskip
Table 1. Contribution of fluctuation modes to partition functions
\end{center}
\end{small}
As can be immediately seen, the products of relevant factors
do agree in the two formalisms, so to complete our proof
it only remains to explain individual factors.
Basically, the only factors which we have not explained
are those for ghosts.

Recall that the fluctuation action for the pure spinor ghosts
to the quadratic order is
simply \(\int(w_{\alpha}D_{-}\lambda^{\alpha} + \widehat{w}_{\hat{\alpha}}D_{+}\widehat{\lambda}^{\hat{\alpha}})\)
so it can be diagonalized just as fermionic fluctuations
by using the matrices \(U\) and \(V\) of~(\ref{eqn:matUV}).
Then, the pure spinor ghosts and their conjugates become \((11+11)\times2\)
left and right moving fields
so their contributions to the partition function combine into \((\det\Box)^{-22}\) as claimed.
Here, \(\Box=4\partial_{+}\partial_{-}\) is the massless Klein-Gordon operator.
Similarly, the reparameterization ghosts in the Green-Schwarz formalism consists of
\(2\) left movers \((b,c)\) and \(2\) right movers \((\widebar{b},\widebar{c})\)
as usual so they contribute \((\det\Box)^{2}\).

Contributions from the fermionic coordinates \((\theta^{\alpha},\widehat{\theta}^{\hat{\alpha}})\) can be inferred from
the analysis of the previous subsection.
In the Green-Schwarz formalism, only a half of \((\theta^{\alpha},\widehat{\theta}^{\hat{\alpha}})\) are propagating
because of the kappa symmetry, and their partition function can be written as
\((\det\nabla_{\text{GS}})\) where \(\nabla_{\text{GS}}\) is defined in~(\ref{eqn:GSeqS}).
In the pure spinor formalism, partition function of \((\theta^{\alpha},\widehat{\theta}^{\hat{\alpha}})\)
can be written as \((\det\nabla_{\text{GS}})(\det\Box)^{24}\)
and is interpreted as coming from Green-Schwarz like degrees of freedom \((S^{A},\widehat{S}^{A})\)
and the rest consiting of \((8+16)\times 2\) left and right moving variables
\((\widehat{T}^{\dot{A}},\widehat{\Psi}^{\hat{\alpha}})\) and \((T^{\dot{A}},\Psi^{\alpha})\).
Actual computation of \((\det\nabla_{\text{GS}})\) is not necessarily easy,
but the difficulty does not hamper the comparison of
the Green-Schwarz and pure spinor formalisms.

As for bosonic fluctuations \((\tilde{t},\tilde{x}^{i},\tilde{\psi},\tilde{\phi})\),
(\(i=1,\dotsc,7\); \(\tilde{t}=t-\underline{t}\) etc.),
recall that they are governed by the same quadratic action~(\ref{eqn:GSS2B}) in the two formalisms,
so the detailed study of their partition functions is not really necessary
for showing the equivalence.
However, it is of some interest to look into their structures.
For a classical solution of the type discussed in this article
partition function factorizes into a product of functional determinants as
\((\det\Delta_{3})^{-1}(\det\Delta_{7})^{-1}\)
where \(\Delta_{3}\) and \(\Delta_{7}\) are some second order matrix differential operators that act on
\((\tilde{t},\tilde{\psi},\tilde{\phi})\in \mathbb{R}_{t}\times S^{2}\)
and on the remaining bosonic fluctuations \(\tilde{x}^{i}\) (\(i=1,\dotsc,7\)).
Actually, the operator \(\Delta_{7}\) is diagonal in the present setting
and \((\det\Delta_{7})=(\det\Box_{\kappa})^{7}\) where \(\Box_{\kappa}\) is the Klein-Gordon operator with mass \(\kappa\).
The other factor \(\Delta_{3}\) acts as \(\Box\) on \(\tilde{t}\) and does not mix it with \((\tilde{\psi},\tilde{\phi})\),
but its action on \((\tilde{\psi},\tilde{\phi})\) is complicated in general.
Nevertheless,
if one believes in the equivalence of the conformal gauge computation
to a static gauge (\(\tilde{t}=\tilde{\phi}=0\)) one,
the functional determinant of \(\Delta_{3}\) should further factorize as
\((\det\Delta_{3})=(\det\Box)^{2}(\det \Delta_{\psi})\)
where \(\Delta_{\psi}\) is the second order differential operator acting on \(\tilde{\psi}\) in the static gauge.
In connection with this,
note that it has been argued quite convincingly that \((\det\Delta_{3})\) actually can be factorized
in this way when a folded string is spinning rigidly in an \(AdS_{3}\subset AdS_{5}\)
instead of \(\mathbb{R}_{t}\times S^{2}\)~\cite{Beccaria:2010ry}.

Putting everything together, we have learnt that
the one-loop partition functions of Green-Schwarz and pure spinor formalisms
agree for the classical solutions of subsection~\ref{subsec:family},
and that the partition function is given as
\begin{align}
\label{eqn:physicalZ}
Z = (\det\Delta_{\psi})^{-1}(\det\Box_{\kappa})^{-7}(\det\nabla_{\text{GS}})
\mathperiod
\end{align}
Since the one-loop partition function is related to
the one-loop correction \(\Delta E\) to spacetime energy
in the present setup, this amounts to a proof of
the equivalence of \(\Delta E\) computed in the two formalisms.

\paragraph{Fluctuation spectra}

It is tempting to interprete the agreement of the partition functions
as indicating that the pure spinor partition function
receives non-trivial contributions only from
physical fluctuations, i.e.\ from BRST cohomologies.
Such an interpretation is possible if, after a quantization,
one can construct transverse DDF operators~\cite{Del Giudice:1971fp} that generate the BRST cohomologies.
The DDF operators should be in one-to-one correspondence with
the transverse oscillators of the lightcone Green-Schwarz formalism,
and completeness of the DDF operators implies
that the remaining degrees of freedom
form BRST quartets with a BRST trivial Hamiltonian.

Although an explicit quantization of fluctuation is not easy in general
even in the Green-Schwarz formalism, it is straightforward
around a point-like rotating string of Berenstein, Maldacena and Nastase~\cite{Berenstein:2002jq}.
The semiclassical analysis around the BMN string in the pure spinor formalism
is just a linearization of the formalism in
a Ramond-Ramond plane-wave background~\cite{Berkovits:2002zv,Berkovits:2008ga}.
We here wish to explain briefly how a physical state of the lightcone Green-Schwarz formalism
is mapped to a BRST cohomology in this case.

In the plane-wave background, physical states of lightcone Green-Schwarz formalism
are described by \(8\) massive bosonic fields \(x^{I}\) and \(8\) pairs of massive fermionic fields
\((S^{A},\widehat{S}^{A})\), where \(I\) and \(A\) are the vector and chiral spinor of \(SO(4)\times SO(4)\)~\cite{Metsaev:2002re}.
As explained in subsection~\ref{subsec:fermifluc},
it is easy to identify the fields with same properties in the pure spinor formalism
at a linearlized level.
Remaining degrees of freedom are lightcone coordinates \(x^{\pm}\),
extra fermionic coordinates \((T^{A},\widehat{T}^{A})\) and \((\theta^{\dot{A}},\widehat{\theta}^{\dot{A}})\),
and pure spinor ghosts \((w_{\alpha},\lambda^{\alpha},\widehat{w}_{\hat{\alpha}},\widehat{\lambda}^{\hat{\alpha}})\).
Although the modes of \((x^{I},S^{A},\widehat{S}^{A})\)
do not directly generate the BRST cohomology,
it should be able to show that elements in their Fock space are in one-to-one
correspondence with BRST cohomologies at ghost number \((1,1)\)
by adopting the methods of~\cite{Berkovits:2000nn} or~\cite{Aisaka:2004ga}
developed for a flat background.

\section{Conclusion}
\label{sec:conclusion}

In this article we have explained how
the one-loop semiclassical analyses of Green-Schwarz
and pure spinor superstrings in an \(AdS_{5}\times S^{5}\) background
are related.
In particular, we have shown that one-loop corrections to
spacetime energies of a classical solution is the same when the solution
is rigid and contained in an \(\mathbb{R}_{t}\times S^{2}\subset AdS_{5}\times S^{5}\).
We would like to interprete the result as a support for
the equivalence of the two formalisms at a semiclassical level.

Let us recapture the main points:
\begin{enumerate}
\item Any purely bosonic classical solution of the Green-Schwarz
formalism can be regarded as a classical solution of the pure spinor formalism
describing the same classical string.
\item To the quadratic order, actions for bosonic fluctuations around a generic classical solution
are the same for the two formalisms.
(Structures at higher orders are different because of their coupling to fermionic fluctuations.)
By contrast,
quadratic actions for fermionic fluctuations are different, yet their structures are strikingly similar.
See equations~(\ref{eqn:GSAdSaction}) and (\ref{eqn:PSAdSaction1}).
\item When a classical string is rigid and contained in an \(\mathbb{R}_{t}\times S^{2}\subset AdS_{5}\times S^{5}\),
the one-loop correction \(\Delta E\) to its spacetime energy is given by the zero point energy of the worldsheet Hamiltonian \(H_{2}\),
both in Green-Schwarz and pure spinor formalisms.
To show that \(\Delta E\) are the same in two formalisms,
it therefore suffices to show that the one-loop partition functions are the same.
Moreover, in view of the second item, it is enough to compare
the partition functions of fermions and ghosts.

\item Even if the rigidity assumption in the previous item is dropped,
fermionic fluctuations in pure spinor formalisms can be
separated into the Green-Schwarz fermions \((S^{A},\widehat{S}^{A})\)
and the rest consisting of \((8+16)\times 2\) left and right movers.
There is a minor coupling between the two types of degrees of freedom,
but the partition function factorizes to the contributions from the two.
\item
Reparameterization \(bc\) ghosts in Green-Schwarz formalism in a conformal gauge
consists of \((1+1)\times 2\) left and right movers.

Pure spinor ghosts are also massless and consists of \((11+11)\times 2\)
left and right movers.
\item The combined partition function of the extra fermions and ghosts in the pure spinor formalism
coincides with that of the \(bc\) ghosts in the Green-Schwarz formalism.
This shows that the total partition functions of the two formalisms are the same.
Hence, if the string is rigid, the one-loop correction \(\Delta E\) to the spacetime energy
computed in the two formalisms agree.
\end{enumerate}
It is natural to ask how far does the equivalence above can be generalized.
As a matter of fact, we believe that the agreement of one-loop partition functions
holds quite generally.
Indeed, around any classical configuration,
\(D_{-}\theta^{\alpha}=D_{+}\widehat{\theta}^{\hat{\alpha}}=0\) gives
a solution to fluctuation equations of motion for the pure spinor formalism
and the equations of motion for the combination
\((\Theta^{\alpha},\widehat{\Theta}^{\hat{\alpha}})\equiv(D_{-}\theta^{\alpha},D_{+}\widehat{\theta}^{\hat{\alpha}})\)
is closely related to that of Green-Schwarz formalism.
Decoupling between the \(D_{-}\theta^{\alpha}=D_{+}\widehat{\theta}^{\hat{\alpha}}=0\) sector
and the \((\Theta^{\alpha},\widehat{\Theta}^{\hat{\alpha}})\) sector,
and splitting of \((\Theta^{\alpha},\widehat{\Theta}^{\hat{\alpha}})\)
into the Green-Schwarz \((S^{A},\widehat{S}^{A})\) variables and the rest
depend on some details of the classical solution in concern,
but it appears reasonable to expect that the combined partition function
of fermions and ghosts in the pure spinor formalism just gives
\((\det\nabla_{\text{GS}})(\det\Box_{bc})^{2}\)
whenever the Green-Schwarz partition function factorizes as in table~1.
It would be interesting to explicitly check these expectations
by studying classical strings in
\(\mathbb{R}_{t}\times S^{3}\in AdS_{5} \times S^{5}\) and \(AdS_{3}\times S^{1}\in AdS_{5}\times S^{5}\)
(so-called \(SU(2)\) and \(SL(2)\) sectors).

The interpretation of the agreement of partition functions
requires additional consideration.
In this article, to obtain a simple relation between
\(\Delta E\) and the worldsheet Hamiltonian \(H_{2}\),
we have put a rigidity assumption on our strings in \(\mathbb{R}_{t}\times S^{2}\).
Presumably, the simple relation continues to hold
as long as \(\underline{t}=\kappa\tau\) (targetspace time is proportional to worldsheet time classically)
and the classical motion is periodic in time.
However, a direct proof purely within a conformal gauge is not necessarily easy.

Extension along another obvious direction, namely,
comparison of semiclassical Green-Schwarz and pure spinor formalisms at two-loops
and higher is of course important.
At higher loops, structures of bosonic fluctuations in the two formalisms
are no longer the same due to their coupling to fermions (and ghosts).
Also, quartic self-coupling of ghosts \(N\widehat{N}\),
which is essential for the conformal invariance of the model,
should play an important role to establish an equivalence.
It would be interesting to understand the relation explicitly.

True power of the pure spinor formalism, however, should be in its generality. 
The fact that one may treat all classical solutions uniformly
without being bothered with gauge fixing
appears to make it more suitable for exploiting integrability.
As is well known, both Green-Schwarz and pure spinor superstrings
in the \(AdS_{5}\times S^{5}\) background possess
Lax connections whose flatness imply classical equations of motion~\cite{Bena:2003wd}\cite{Vallilo:2003nx}.
To show the integrals of motion generated by the flat connection
to be mutually commutative,
one wishes to check that the connection satisfies
a certain exchange algebra introduced by Maillet~\cite{Maillet:1985ek}.
In~\cite{Magro:2008dv}\cite{Vicedo:2009sn},
the Green-Schwarz flat connection of Bena-Polchinski-Roiban~\cite{Bena:2003wd}
have been investigated within the Dirac-Hamiltonian formalism,
and it have been found that the flat connection
have to be improved by adding phase space constraints
to satisfy the exchange property.
Moreover, the flat connection after the improvement
have been found to be the one in the pure spinor formalism constructed by one of the authors~\cite{Vallilo:2003nx}
(minus the ghost contribution).
This indicates that the pure spinor formalism is a properly gauge fixed
version of the Green-Schwarz formalism.
It would be reasonable and interesting, therefore,
to exploit the integrability of the pure spinor formalism systematically.

Ultimately, one would like to solve the string theory in the \(AdS_{5}\times S^{5}\)
background by an exact quantization.
In supercoset models describing Ramond-Ramond backgrounds,
currents \(J\) are not holomorphic unlike in the Wess-Zumino-Witten models,
so their operator product expansions are difficult to control.
It is just about hopeless to find a good theory
for arbitrary non-holomorphic currents, but one
could hope that the Ramond-Ramond supercoset models form
a good class of conformal field theories.
For example, the currents \(J\) are actually
covariantly holomorphic as in~(\ref{eqn:PSclassEOM})
indicating an enhancement of chiral algebra from the Virasoro algebra~\cite{Bershadsky:1999hk}.

We would like to come back to some of these issues in the near future.

\section*{Acknowledgments}

We would like to thank Nathan Berkovits and Volker Schomerus
for useful discussions, encouragement and comments on a draft of this article,
and IFT/UNESP where a part of this research was done.
YA would also like to thank Carlo Meneghelli and Beno\^{i}t Vicedo
for useful discussions and encouragement,
and is grateful to Soo-Jong Rey for a discussion on Goldstone modes
in the semiclassical pure spinor formalism that initiated our study.
LIB would like to thank Victor O.\ Rivelles for discussions.
The work of YA is supported by SFB~676
and a part of the work was done under a support of FAPESP grant 06/59970-5.
The work of LIB was partially supported by
Conselho Nacional de Desenvolvimento Cient\'{i}fico e Tecnol\'{o}gico (CNPq)
and Funda\c{c}\~{a}o de Apoio \`{a}
Pesquisa do Estado do Rio Grande do Norte (FAPERN).
The work of BCV is partially supported by FONDECYT grant number 1120263.

\appendix
\section{Appendix: Notation and conventions}
\label{sec:appendix}

\subsection*{Worldsheet}

Worldsheet of a string is assumed to be a cylinder.
We keep the worldsheet to be Minkowskian
except in subsection~\ref{subsec:revflat} where we review
the pure spinor formalism in a flat background.
\begin{itemize}
\item Coordinates on worldsheet cylinder:
\begin{align}
\sigma^{\mu} & =(\tau,\sigma),\quad \sigma+2\pi = \sigma,\quad (\eta^{\tau\tau}=-1,\; \eta^{\sigma\sigma}=1)
\end{align}
\item Lightcone:
\begin{align}
\sigma^{\pm} = \tau\pm \sigma,\quad \partial_{\pm} = {1\over2}(\partial_{\tau}\pm\partial_{\sigma})
\end{align}
\item Coordinates on a (Euclidean) complex plane:
\begin{align}
z = \mathe^{\tilde{\tau} + \mathi \sigma},\quad \widebar{z} = \mathe^{\tilde{\tau} - \mathi \sigma},\quad (\tilde{\tau} = \mathi \tau)
\end{align}
\end{itemize}

\subsection*{Gamma matrices}

\begin{itemize}
\item \(SO(9,1)\) and \(SO(4,1)\times SO(5)\) gamma matrices of size \(16\times16\)
are denoted by \((\gamma_{a})_{\alpha\beta}\) and \((\gamma_{a})^{\alpha\beta}\).
They satisfy \(\{(\gamma_{a})_{\alpha\beta},(\gamma_{b})^{\beta\gamma}\} = 2\eta_{ab}\delta_{a}{}^{\gamma}\).
We assume that a basis for spinors is chosen so that \((\gamma_{0})_{\alpha\beta} = -(\gamma_{0})^{\alpha\beta} = 1_{16}\).
\item For \(SO(4,1)\times SO(5)\) there is an invariant tensor given by an antisymmetric product
\(\gamma^{01234}\) of gamma matrices:
\begin{align}
\eta_{\hat{\alpha}\alpha} &\equiv -\eta_{\alpha\hat{\alpha}} \equiv (\gamma^{01234})_{\hat{\alpha}\alpha},\quad
\eta^{\hat{\alpha}\alpha} \equiv -\eta^{\alpha\hat{\alpha}} \equiv (\gamma_{01234})^{\hat{\alpha}\alpha} \\
\eta_{\alpha\hat{\alpha}}\eta^{\hat{\alpha}\beta} &= \delta_{\alpha}{}^{\beta},\quad
\eta_{\hat{\alpha}\alpha}\eta^{\alpha\hat{\beta}} = \delta_{\hat{\alpha}}{}^{\hat{\beta}}
\end{align}
We use \(\eta\) to define spinor indices with hats.
In particular, gamma matrices with hatted indices are defined via
\begin{align}
(\gamma_{a})^{\hat{\alpha}\hat{\beta}} = \eta^{\hat{\alpha}\alpha}(\gamma_{a})_{\alpha\beta}\eta^{\beta\hat{\beta}},\quad
(\gamma_{a})_{\hat{\alpha}\hat{\beta}} = \eta_{\hat{\alpha}\alpha}(\gamma_{a})^{\alpha\beta}\eta_{\beta\hat{\beta}}
\end{align}
\item In the context of \(\mathfrak{psu}(2,2|4)\),
\(\eta_{\alpha\hat{\alpha}}\) can be identified as the ``spinor metric''
coupling \(\mathfrak{g}^{1}\) and \(\mathfrak{g}^{3}\).
See below.
\end{itemize}

\subsection*{\boldmath $\mathfrak{psu}(2,2|4)$}

\begin{itemize}
\item Generators:
\begin{align}
T_{A} = (T_{a},T_{ab};T_{\alpha},T_{\hat{\alpha}}) = (P_{a},L_{ab};Q_{\alpha},\widehat{Q}_{\hat{\alpha}}),\quad
A = (a,ab;\alpha,\hat{\alpha})
\end{align}
\item $\mathbb{Z}_{4}$ structure:
\begin{align}
\mathfrak{psu}(2,2|4)
&= \mathfrak{g}^{0}+\mathfrak{g}^{1}+\mathfrak{g}^{2}+\mathfrak{g}^{3} \\
L_{ab} \in \mathfrak{g}^{0} &,\quad
P_{a} \in \mathfrak{g}^{2},\quad
Q_{\alpha} \in \mathfrak{g}^{1},\quad
\widehat{Q}_{\hat{\alpha}} \in \mathfrak{g}^{3}
\end{align}
Both commutation relations and inner product below respect $\mathbb{Z}_{4}$:
\begin{align}
[\mathfrak{g}^{i},\mathfrak{g}^{j}] = \mathfrak{g}^{i+j},\quad
\str( \mathfrak{g}^{i}\mathfrak{g}^{j} ) \ne 0\;\;
\text{only when $i+j=0$}
\end{align}
\item Trace metric:
\begin{align}
\eta_{AB} &\equiv \str(T_{A}T_{B}),\quad \\
&\str(P_{a}P_{b}) = \eta_{ab} \\
&\str(L_{ab}L_{cd}) = -R_{abcd}
{} = \begin{cases}
{} -(\eta_{ac}\eta_{bd} - \eta_{bc}\eta_{ad}) & AdS_{5}\\
{} \delta_{ac}\delta_{bd} - \delta_{bc}\delta_{ad} & S^{5}
\end{cases} \\
&\str(\widehat{Q}_{\hat{\alpha}}{Q}_{\beta}) = -\str({Q}_{\beta}\widehat{Q}_{\hat{\alpha}})
{} = \gamma^{01234}_{\hat{\alpha}\beta}
\end{align}
\item Commutation relations: \\
It is convenient to split \(a=(a'',a')\) where
\(a''=0,\dotsc,4\) are \(AdS_{5}\) directions
and \(a'=5,\dotsc,9\) are \(S^{5}\) directions.
Non-trivial commutation relations are then
\begin{align}
[P_{aa''},P_{b''}] &= L_{a''b''},\quad
[P_{a'},P_{b'}] = -L_{a'b'}
\\
[L_{a''b''},P_{c''}] &= \eta_{b''c''} P_{a''} - \eta_{a''c''}P_{b''},\quad
[L_{a'b'},P_{c'}] = \eta_{b'c'} P_{a'} - \eta_{a'c'}P_{b'}
\\
[L_{a''b''},L_{c''d''}] &= \eta_{b''c''}L_{a''d''} \pm \cdots,\quad
[L_{a'b'},L_{c'd'}] = \eta_{b'c'}L_{a'd'}\pm\cdots
\\
[L_{ab},Q_{\alpha}] &= -{1\over2}({\gamma}_{ab})_{\alpha}{}^{\beta}Q_{\beta},\quad
[L_{ab},\widehat{Q}_{\hat{\alpha}}]
{} = -{1\over2}({\gamma}_{ab})_{\hat{\alpha}}{}^{\hat\beta}\widehat{Q}_{\hat{\beta}}
\\
[P_{a},Q_{\alpha}] &= {1\over2}(\eta\gamma_{a})_{\alpha}{}^{\hat{\beta}}\widehat{Q}_{\hat{\beta}},\quad
[P_{a},\widehat{Q}_{\hat{\alpha}}] = -{1\over2}(\eta\gamma_{a})_{\hat{\alpha}}{}^{\beta}Q_{\beta}
\\
\{Q_{\alpha},Q_{\beta}\} &= \gamma^{a}_{\alpha\beta}P_{a},\quad
\{\widehat{Q}_{\hat{\alpha}},\widehat{Q}_{\hat{\beta}}\} = \gamma^{a}_{\hat{\alpha}\hat{\beta}}P_{a}
\\
\{Q_{\alpha},Q_{\hat{\beta}}\}
&= {1\over2}(\eta\gamma^{a''b''})_{\alpha \hat{\beta}}L_{a''b''}
{} - {1\over2}(\eta\gamma^{a'b'})_{\alpha \hat{\beta}}L_{a'b'}
\end{align}

\end{itemize}


\end{document}